\pdfoutput=1
\def\RepoUrl/{https://github.com/Kasra-Z-A/yield_paper}
\pdfoutput=1 
\documentclass[a4paper,UKenglish,cleveref,autoref,thm-restate]{lipics-v2021}

\ifdefined\anonymousAck\else\hideLIPIcs\fi

\usepackage{siunitx}
\usepackage{multirow}
\usepackage{tablefootnote}
\usepackage{booktabs}
\usepackage{array}
\usepackage{threeparttable}
\usepackage{subcaption}
\usepackage{tikz}
\usepackage{pifont}
\usetikzlibrary{shapes.geometric}

\newcolumntype{L}[1]{>{\raggedright\arraybackslash}p{#1}}
\newcolumntype{C}[1]{>{\centering\arraybackslash}p{#1}}
\newcolumntype{R}[1]{>{\raggedleft\arraybackslash}p{#1}}

\crefname{definition}{Definition}{Definitions}
\crefname{proposition}{Proposition}{Propositions}
\crefname{theorem}{Theorem}{Theorems}
\crefname{corollary}{Corollary}{Corollaries}
\crefname{example}{Example}{Examples}
\crefname{remark}{Remark}{Remarks}

\def\YearnDeposits/{\num{15663293}}
\def\YearnWithdrawals/{\num{10905492}}
\def\YearnInvestments/{\num{13624423}}
\def\CianDeposits/{\num{53996952}}
\def\CianWithdrawals/{\num{17515202}}
\def\CianInvestments/{\num{53882811}}

\def\YearnDepositsRounded/{\num{15.7}M}
\def\YearnWithdrawalsRounded/{\num{10.9}M}
\def\YearnInvestmentsRounded/{\num{13.6}M}
\def\CianDepositsRounded/{\num{54.0}M}
\def\CianWithdrawalsRounded/{\num{17.5}M}
\def\CianInvestmentsRounded/{\num{53.9}M}

\def\YearnDepositsCount/{\num{986}}
\def\YearnWithdrawalsCount/{\num{886}}
\def\YearnInvestmentsCount/{\num{738}}
\def\CianDepositsCount/{\num{171}}
\def\CianWithdrawalsCount/{\num{39}}
\def\CianInvestmentsCount/{\num{14}}

\def\YearnTransactionsTotal/{\num{228443}}
\def\YearnTransactionsUnique/{\num{2459}}
\def\YearnTokenTransfers/{\num{5575}}
\def\YearnYield/{\num{5.41}\%}

\def\CianTransactionsTotal/{\num{152125}}
\def\CianTransactionsUnique/{\num{921}}
\def\CianTokenTransfers/{\num{1963}}
\def\CianYield/{\num{4.22}\%}

\def\YearDepositToInvestRate/{\num{86,9}}
\def\CianDepositToInvestRate/{\num{99,7}}

\def\YearnMaxDeposit/{\num{1.1}M}
\def\CianMaxDeposit/{\num{10.1}M}

\def\YearnDepositMean/{\num{24}k}
\def\YearnDepositMedian/{\num{0.5}k}
\def\CianDepositMean/{\num{340}k}
\def\CianDepositMedian/{\num{3}k}

\def\FedFundsRate/{\num{4.77}\%}
\def\STETHBaseline/{\num{2.94}\%}

\def\CianLeverageMultiplier/{\num{5}}

\def\CianFlashloanSeed/{\num{2.48}M}
\def\CianFlashloanAmount/{\num{9.87}M}
\def\CianCollateralAfterLeverage/{\num{12.35}M}

\ifdefined\RepoUrl\else
  \def\RepoUrl/{https://anonymous.4open.science/r/yield_paper-D840}
\fi

\title{DeFi Yield Aggregators: Analysing Investment Strategies and Structural Dependencies}
\titlerunning{DeFi Yield Aggregators}

\author{Stefan Kitzler\footnote{These authors contributed equally to this work.}}{Complexity Science Hub \& AIT Austrian Institute of Technology, Austria}{kitzler@csh.ac.at}{}{}

\author{Kasra Zarinehbaf Asadi\footnotemark[1]}{Complexity Science Hub \& TU Wien, Austria}{asadi@csh.ac.at}{}{}

\author{Svetlana Kremer}{Complexity Science Hub, Austria}{kremer@csh.ac.at}{}{}

\author{Bernhard Haslhofer}{Complexity Science Hub, Austria}{haslhofer@csh.ac.at}{}{}

\authorrunning{S. Kitzler, K. Zarinehbaf Asadi, S. Kremer, and B. Haslhofer}

\Copyright{Stefan Kitzler, Kasra Zarinehbaf Asadi, Svetlana Kremer, and Bernhard Haslhofer}



\category{}
\relatedversion{}
\supplement{}
\funding{}

\ifdefined\anonymousAck
\acknowledgements{Acknowledgements anonymised for double-blind review. \vspace{1\baselineskip}}
\else
\acknowledgements{This work is partially funded by the Austrian security research program KIRAS of the Federal Ministry of Finance (BMF) under the project LLEA (grant agreement 926183). \vspace{1\baselineskip}}
\fi

\EventEditors{Aggelos Kiayias and Maria Kyropoulou}
\EventNoEds{2}
\EventLongTitle{8th Conference on Advances in Financial Technologies (AFT 2026)}
\EventShortTitle{AFT 2026}
\EventAcronym{AFT}
\EventYear{2026}
\EventDate{October 6--9, 2026}
\EventLocation{London, United Kingdom}
\EventLogo{}
\SeriesVolume{8}
\ArticleNo{1}

\ifdefined\anonymousAck
\hypersetup{pdfauthor={},pdftitle={},pdfsubject={},pdfkeywords={},pdfcreator={},pdfproducer={}}
\fi

\begin{document}

\ifdefined\anonymousAck\else\nolinenumbers\fi

\maketitle

\ifdefined\anonymousAck
\vspace{2\baselineskip}
\else
\vspace{1\baselineskip}
\fi

\begin{abstract}
Yield aggregators are financial services in Decentralised Finance (DeFi) providing automated investment management and return optimisation for users.
In this study, we investigate the operational mechanisms and monetary flows of two major yield aggregators, Yearn Finance and Cian, over the period from May 4, 2024 to May 3, 2025.
Our supporting conceptual framework decomposes yield aggregator operations into user investment and strategy management cycles.
Using a network approach for \YearnTransactionsUnique/ Yearn and \CianTransactionsUnique/ Cian transactions, we trace protocol interactions and capital flows across the ecosystem.
Users invested \YearnDepositsRounded/ USD into Yearn's USDC vault, which generated yield through liquidity provision and dynamic allocation across DeFi protocols.
Cian, deployed later, attracted \CianDepositsRounded/ USD into its staked-ETH (stETH) vault and implemented sophisticated leverage through flashloan-enabled recursive staking.
Yearn's USDC vault achieves an annual yield of \YearnYield/, while Cian's stETH vault produces \CianYield/ with higher risk exposure.
We use the operational insights from our analysis to extend the existing DeFi Stack Reference Model (DSR) with new financial primitives to highlight structural risk dependencies.
Overall, our findings show that strategic complexity in yield aggregation does not necessarily translate into higher returns but materially expands risk exposure.
\end{abstract}

\ifdefined\anonymousAck
\vspace{2\baselineskip}
\fi

\setcounter{totalnumber}{10}
\setcounter{topnumber}{10}
\setcounter{bottomnumber}{10}
\renewcommand{\topfraction}{.99}
\renewcommand{\bottomfraction}{.99}
\renewcommand{\textfraction}{.01}

\section{Introduction}\label{sec:intro}

Decentralised finance (DeFi) continues to evolve through the emergence of new protocol types, sophisticated financial primitives, and increasingly interconnected services. A major catalyst for recent development has been Ethereum’s transition from proof-of-work (PoW) to proof-of-stake (PoS), which accelerated innovation across the ecosystem and fundamentally expanded the architecture of DeFi. These developments have enabled new forms of yield generation, capital-efficient investment strategies, and increasingly complex financial interactions that tightly couple protocols, assets, liquidity pools, and infrastructure layers across the ecosystem.

Despite this growing structural complexity, existing literature has largely focused on individual DeFi primitives, such as decentralised exchanges~\cite{Xu:2021}, lending platforms~\cite{Bartoletti:2020}, or staking protocols~\cite{Gogol2024empirical,Gogol2024sok}. Consequently, higher-layer protocols that actively compose these primitives remain comparatively underexplored. Specifically, there is a lack of empirical research regarding how these aggregators structure capital allocation, embed inter-protocol dependencies, and facilitate cross-protocol risk propagation. This constitutes a significant gap in our understanding of how composability translates into both yield generation and systemic exposure in practice.

Yield aggregators provide an optimal lens to address this gap, as they explicitly leverage composability by integrating underlying DeFi protocols into unified yield optimisation strategies. Their internal operational mechanics offer a natural empirical setting to investigate how structural dependencies are formed and sustained across the DeFi stack, as well as how capital flows and associated risks are orchestrated across different protocol layers.

Motivated by this perspective, this paper investigates the following research question:
\textit{What protocol dependencies underpin the profit optimisation strategies of DeFi yield aggregators?}
We address this question through an empirical analysis of two yield aggregators, leveraging methods from network and data science to trace token flows and to reconstruct protocol interactions and internal operational logic. Through this analysis, we aim to characterise the dependency structures embedded in yield generation strategies and contribute to a more grounded understanding of how multi-layered DeFi protocols operate in practice, including the channels through which risk propagates across the ecosystem.

Our key contributions and results are summarised as follows:
\begin{enumerate}
    \item We conduct a data-driven analysis of two yield aggregators, Yearn Finance and Cian, analysing yield generation and investment strategies over the period from May 4, 2024 to May 3, 2025.
    We trace protocol interactions and monetary flows, examining \YearnTokenTransfers/ token transfers of Yearn Finance and \CianTokenTransfers/ transfers for Cian.

    \item We conceive a framework that decomposes yield aggregator operations into user investment and strategy management cycles, providing a structured foundation for understanding of these platforms.
    During the observation period, users invested \YearnDeposits/ (\YearnDepositsRounded/) USD into Yearn's USDC vault and \CianDeposits/ (\CianDepositsRounded/) USD into Cian's stETH vault,
    with \YearDepositToInvestRate/\% and \CianDepositToInvestRate/\% allocated to active yield generation, respectively.

    \item We identify the primary DeFi protocols used for yield generation,
    including lending platforms and restaking protocols.
    We reconstruct and compare investment strategies, revealing distinct approaches:
    Yearn primarily employs liquidity provision, yielding \YearnYield/ annually on USDC,
    while Cian implements a sophisticated leverage scheme with multiple yield sources,
    achieving \CianYield/ annually on ETH.
    We further examine Cian's yield generation mechanisms
    by analysing its use of flash loans to amplify leverage on user deposits within Aave.

    \item Based on our analysis, we map yield
     aggregators components and associated risks to the DeFi architectural layers to understand structural dependencies and potential risk contagion pathways.
\end{enumerate}

The rest of the paper is organised as follows. Section~\ref{sec:bg} provides a conceptual architecture of yield generation and reviews related work. Section~\ref{sec:data} describes the data collection and processing pipeline. Our analysis and key results are presented in Section~\ref{sec:analysis}. Section~\ref{sec:disc} compares the analysed yield aggregator protocols, outlines risk associations, and discusses our findings. Section~\ref{sec:concl} concludes the paper.

\section{Background}\label{sec:bg}

\subsection{Architecture of Yield Aggregators}\label{subsec:ya-arch}
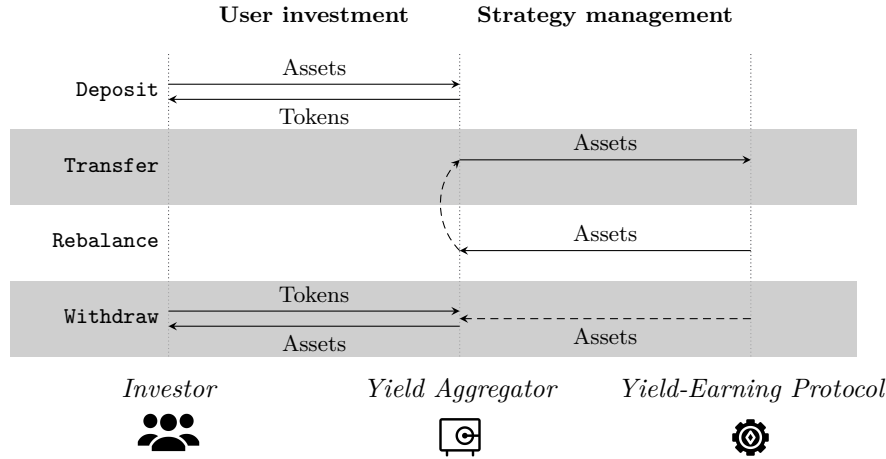
\begin{figure}[!htb]
\centering
\begin{tikzpicture}[>=stealth, x=0.7cm, font=\footnotesize]
   \node[] at (2.75,4.5) {\textbf{User investment\vphantom{Aq}}};
   \node[] at (8.25,4.5) {\textbf{Strategy management\vphantom{Aq}}};
   \foreach \x in {0,5.5,11} {
        \draw[densely dotted, darkgray] (\x,0) -- (\x,4);
    }
    \foreach \y in {0,2} {
        \fill[lightgray,opacity=.75] (-3,\y) rectangle (13,\y+1);
    }
    \begin{scope}[font=\itshape]
        \node[below=5pt,align=center] at (0,0) {Investor\\[5pt]\includegraphics[height=.5cm]{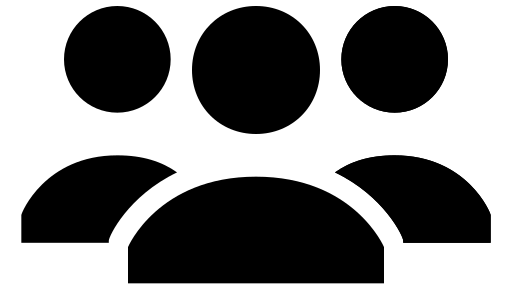}};
        \node[below=5pt,align=center] at (5.5,0) {Yield Aggregator\\[5pt]\includegraphics[height=.5cm]{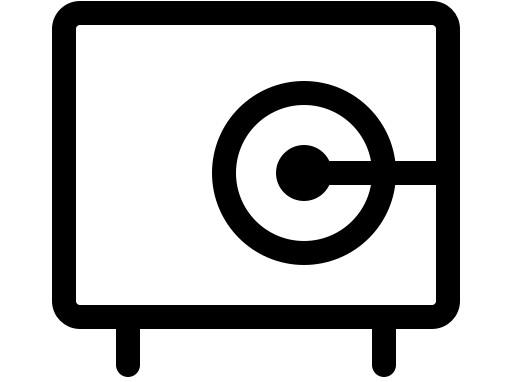}};
        \node[below=5pt,align=center] at (11,0) {Yield-Earning Protocol\\[5pt]\includegraphics[height=.5cm]{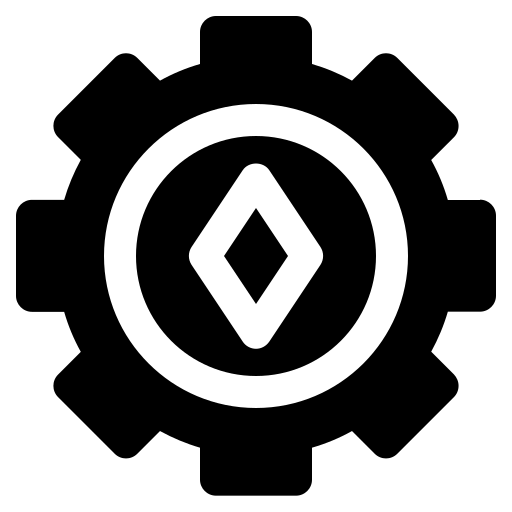}};
    \end{scope}
    \begin{scope}[]
        \node[left] at (0,3.5) {\texttt{Deposit\vphantom{Aq}}};
        \node[left] at (0,2.5) {\texttt{Transfer\vphantom{Aq}}};
        \node[left] at (0,1.5) {\texttt{Rebalance\vphantom{Aq}}};
        \node[left] at (0,0.5) {\texttt{Withdraw\vphantom{Aq}}};
    \end{scope}

    \draw[->] (0,3.6)--(5.5,3.6) node[pos=.5,above]{\footnotesize Assets};
    \draw[->] (5.5,3.4)--(0,3.4) node[pos=.5,below]{\footnotesize Tokens};
    \draw[->] (5.5,2.6)--(11,2.6) node[pos=.5,above]{\footnotesize Assets};
    \draw[->] (11,1.4)--(5.5,1.4) node[pos=.5,above]{\footnotesize Assets};
    \draw[->, bend left=45, densely dashed] (5.5,1.4) to (5.5,2.6);
    \draw[->] (0,.6)--(5.5,.6) node[pos=.5,above]{\footnotesize Tokens};
    \draw[->] (5.5,.4)--(0,.4) node[pos=.5,below]{\footnotesize Assets};
    \draw[->,densely dashed] (11,.5)--(5.5,.5) node[pos=.5,below]{\footnotesize Assets};
\end{tikzpicture}
    \caption{Simplified overview of yield aggregator operations }
    \label{fig:ya-cycles}
    \end{figure}

The operational architecture of DeFi yield aggregators can be analysed by identifying the key actors involved in the yield generation process and their interactions (see Fig.~\ref{fig:ya-cycles}).

In a typical setup, the following actors can be distinguished:
\begin{itemize}
    \item \textbf{Investors (I)}: individuals or entities who deposit assets to earn yield. Investors can be categorised as either externally owned accounts (EOAs) or contract accounts (CAs). EOAs are typically controlled by individuals, whereas CAs are smart contracts that represent other protocols or automated strategies.
    \item \textbf{Yield-Earning Protocol (P)}: an external DeFi protocol that generates yield in exchange for some financial activity. Common examples include decentralised exchanges (DEXs)~\cite{Xu:2021}, lending~\cite{Bartoletti:2020}, and (liquid) (re-)staking protocols~\cite{Gogol2024empirical,Gogol2024sok}. DEXs and lending protocols allow liquidity providers to deposit funds into pools and earn rewards from fees paid by users of these services. Liquid staking and restaking protocols issue tradable tokens that represent assets invested to participate in a blockchain's consensus, continuously accruing the corresponding rewards. Ethereum-based examples of yield-earning protocols include Aave and Compound (lending), the Morpho optimiser built on top of them, Lido (liquid staking), EigenLayer (restaking), and Renzo (liquid restaking).
    \item \textbf{Yield Aggregator (YA)}: a protocol that automates the management of user deposits by selecting and reallocating funds across yield-earning protocols to maximise returns. The most popular Ethereum-based YAs are Yearn Finance, CIAN Yield Layer, Fluid Lite, and Beefy.
\end{itemize}

This structure enables investors to benefit from automated yield optimisation without manually managing positions across multiple underlying protocols, while YAs dynamically orchestrate asset flows and execute investment strategies across the ecosystem. Building on this, the operations of a yield aggregator can be conceptualized as two distinct, yet interrelated cycles: the \textit{user investment cycle} and the \textit{strategy management cycle}.

The \textbf{user investment cycle} captures interactions between investors and the YA, primarily through deposits and withdrawals of assets (see Fig.~\ref{fig:ya-cycles}). An investor deposits assets into a yield aggregator's vault and receives tokens in return, representing a proportional share of the vault's underlying assets. To withdraw, the investor returns these tokens, prompting the YA to automatically source the corresponding assets from one or more protocols when the vault's available liquidity is insufficient. Through these actions, investors steer the aggregator's operations and capital deployment.

The \textbf{strategy management cycle} represents the internal decision-making and execution processes within the YA. It involves initiating investment strategies by allocating aggregated user funds to the chosen yield-earning protocols and re-balancing assets across them in response to changing market conditions (see Fig.~\ref{fig:ya-cycles}). Together, these two cycles offer a simplified view of a yield aggregator's operation, highlighting both the external interactions with investors and the internal mechanisms that generate and manage yield.

\subsection{Yield-Earning Protocols}\label{sec:yield-protocols}

Yield aggregators draw returns from yield-earning DeFi protocols, such as lending, staking, liquid staking, and restaking. We briefly review the relevant concepts of DeFi yield generation.

\textbf{Lending} protocols (e.g., Aave~\cite{aave}, Compound~\cite{compound}) operate as over-collateralised money markets in which lenders deposit assets into a pool from which borrowers can draw against collateral~\cite{Bartoletti:2020}. Interest rates accrue algorithmically based on the availability of liquidity in the pool, with borrowers paying interest and lenders receiving a share of it as yield. In exchange for a deposit, the protocol mints an interest-bearing \emph{receipt token} (e.g., Aave's aTokens) that represents a tradable claim. When the receipt token is returned, the lender receives back the underlying deposit plus the accrued interest. To protect lenders against bad debt~\cite{Perez2021}, active users can spot under-collateralised borrowing positions and liquidate them~\cite{Qin2021a} by repaying part of the outstanding debt in exchange for a discounted share of the borrower's collateral.

\textbf{Staking} rewards participants for locking native tokens in a Proof-of-Stake (PoS) network, where validators propose and attest blocks and, in return, earn issuance and transaction-fee rewards. For misbehaviour, such as double-signing or prolonged downtime, validators are penalised (\textit{slashed}), losing part of their staked tokens. Ethereum transitioned from Proof-of-Work to PoS on September 15, 2022 (``The Merge''), turning staking-based yield into a foundational DeFi service of the Ethereum ecosystem. However, operating a validator on Ethereum requires a fixed deposit of 32~ETH and dedicated infrastructure, which puts solo staking out of reach for most users.

\textbf{Liquid staking} protocols (e.g., Lido, Rocket Pool) give ordinary participants access to staking yield without meeting the validator threshold. In exchange for a deposit, the protocol pools funds across validators and issues a yield-bearing \emph{liquid staking token} (LST), such as stETH or rETH~\cite{alexander2024leveraged,Gogol2024sok}, a receipt token that represents a tradable claim on the underlying stake and its validator rewards.

\textbf{Restaking} platforms allow already-staked ETH, supplied via LSTs, to be redeployed to generate a second yield stream by securing additional services, called \emph{actively validated services} (AVSs). EigenLayer is the dominant restaking platform~\cite{eigenlayer}, with Symbiotic and Karak as competitors. The restaking platform offers the deposited assets as economic security to AVSs (e.g., bridges, oracles, data-availability layers) and, acting as intermediary, slashes the underlying stake if AVS conditions are violated. Restaking platforms themselves do not issue a tradable receipt token to depositors.

\textbf{Liquid restaking} services (e.g. Renzo~\cite{renzo}, Kelp, and ether.fi) close the gap of tradeable receipts by acting as intermediaries between users and the restaking platforms. In exchange for a deposit, they commit user funds to a restaking protocol (predominantly EigenLayer) and issue a yield-bearing receipt token -- the \emph{liquid restaking token} (LRT), such as ezETH, rsETH, or eETH~\cite{alexander2024leveraged} -- that represents the depositor's share of the restaked position.

\textbf{Leveraged lending} is a well-known DeFi technique~\cite{Wang2022} for amplifying yield: an asset is supplied as collateral, the borrowed funds are converted back into the same (or a correlated) asset, and the position is re-deposited recursively. This process is also called \emph{folding} or, more commonly, \emph{looping}~\cite{alexander2024leveraged}, and YAs exploit this in their strategies on top of the primitives above.

\subsection{Related Work}
DeFi yield aggregators have received little academic attention so far. The most relevant works are those by Cousaert et al.~\cite{Cousaert2022} and Xu et al.~\cite{Xu2022b}, who provide the first systematic examinations of yield aggregator protocols. These studies map the design space of yield strategies, characterise common architectural components, and analyse revenue models and associated risks. 
While they establish an important foundation for understanding the operation of yield aggregators, they operate at a high level of protocol abstraction and provide limited empirical scrutiny of on-chain activity and token transfer networks.

Other strands of research that intersect with yield aggregation include analyses of DeFi composability and  protocol interactions. Prior work~\cite{Kitzler2023c} has examined DeFi transaction networks across protocols and assets and their structural metrics using on-chain Ethereum data. Studies of lending protocols such as Aave, MakerDAO, Compound, and Liquity~\cite{zhang2023blockchain} reveal core–periphery structures dominated by major exchanges, challenging assumptions of decentralisation in the DeFi ecosystem. At the token level, Somin et al.~\cite{somin2018network} analyse a large ERC-20 transfer network and identify power-law degree distributions, while \cite{victor2019measuring} shows many individual token networks. More recent work~\cite{alamsyah2024unraveling} finds that the structural properties of token networks vary substantially across several assets, with centralised exchanges consistently serving as key intermediaries.

Since yield aggregators explicitly depend on other protocols by investing in them, they are also exposed to the underlying risks of these protocols and the wider DeFi ecosystem. The systemic risks and the interconnectedness of the DeFi ecosystem is receiving increasing attention in the literature. Weingartner et al.~\cite{weingartner2023deciphering} classify various DeFi risks found in the literature and provide a visual representation of systemic and unsystemic risks in the form of a risk wheel. Aufiero et al.~\cite{aufiero2025mapping} review systemic risks in DeFi and TradFi and propose a framework that maps core risks along two axes: speed of contagion and scale from local to systemic. Zhang et al.~\cite{zhang2026systemic} propose a network-based framework to measure systemic risk. At the vulnerability level, Arora et al.~\cite{arora2026risk} develop a prioritisation framework for DeFi infrastructure that combines incident frequency, economic severity, and adversarial feasibility across 558 exploit incidents from 2021–2025. Closer to the recursive strategies executed by YAs, Wang et al.~\cite{Wang2022} formalise on-chain leverage in lending protocols and quantify the speculative multipliers achievable through recursive borrowing on generic DeFi assets. Alexander~\cite{alexander2024leveraged} extends this lens to leveraged staking and leveraged staking-restaking, showing that depeg and liquidation risks are substantially larger for leveraged restaking tokens than for leveraged LSTs due to thinner LRT liquidity.

Unlike prior work, our study empirically examines how yield aggregators operate in practice.
We extend DeFi research by analysing the token-transfer networks underpinning both user-facing and behind-the-scenes investment cycles, uncovering operational patterns, dependencies, and structural complexities that individual protocol analysis misses.
We further map yield aggregator components and their associated risks onto the DeFi architectural stack, linking protocol-level operations to the systemic risk dimensions discussed above.

\section{Data and Methods}\label{sec:data}

Here, we describes our data collection methodology, pre-processing procedures, and analytical approach for examining yield aggregator operations on Ethereum.
We outline how we gather transaction data, identify relevant protocol interactions, and construct a network approach to investigate both user investment and internal strategy management cycles.

\subsection{Data Collection}

Our data come from a full Ethereum blockchain node~\cite{erigon} and public DeFi yield aggregator addresses, enriched with Etherscan~\cite{etherscan} labels and 4-byte~\cite{4byte,Sourcify} function signatures.

We select Yearn Finance and Cian, the two largest yield aggregators on Ethereum by total value locked (TVL), according to DeFiLlama~\cite{defillama}.
Beyond market relevance, both represent different maturity stages and pursue distinct investment approaches, offering complementary perspectives on yield aggregation.
For each protocol, we selected one vault and completed a user investment cycle with manual deposit/withdrawal transactions. 
We examined function calls and ground truth vault addresses, cross-checking them with their official websites.
Our collection includes one address for Yearn and eight for Cian.

We run an Ethereum archive node to access \textit{transaction} data (including internal contract calls) and \textit{event logs} from May 4, 2024, to May 3, 2025.
Transactions capture contract interactions for examining usage patterns and investment management.
Event logs, emitted while the transactions are executed, record token transfers and enable us to track fund movements between users and protocols.
We supplement the on-chain data with transfer and protocol logs gathered from Etherscan for practical reasons.

\subsection{Data Pre-processing and Network Construction}

We apply a two-step procedure to curate a dataset of yield aggregator operations,
encompassing contract interactions and related fund movements.

First, we filter for transactions that involve aggregator usage and identify their intent.
We include transactions where at least one call targets a vault address from our collection.
Next, we introduce function signatures~\cite{4byte,Sourcify} to decode the bytecode of called functions.
For each aggregator, we map methods to either user interactions or management operations.

Second, we filter for token transfers that occurred in the selected transactions.
This enables us to establish relationships between the actions taken and the corresponding fund movements.
We also gather address tags from Etherscan~\cite{etherscan} to label protocols involved as counterparts of the transfers.

Building on the pre-processed data, we construct temporal networks where nodes represent addresses (users, vaults, or protocols) and directed edges represent token transfers with timestamps and amounts.
This framework enables time-evolving analysis of usage and investments in DeFi protocols for yield generation.

\subsection{Dataset Summary}

Our dataset contains \YearnTransactionsUnique/ Yearn transactions (\YearnTokenTransfers/ token transfers) and \CianTransactionsUnique/ Cian transactions (\CianTokenTransfers/ token transfers).
The data captures user interactions, including deposit and withdrawal, and internal management operations, including investment for yield generation, during the observation period.
Table~\ref{tab:core_operations_combined} provides an overview of these operations for both protocols, showing the number of transactions (txns) and underlying token transfers.

\begin{table}[!htbp]
    \centering
    \begin{tabular*}{\textwidth}{@{\extracolsep{\fill}}lllrr}
    \hline
    \textbf{} & \textbf{Protocol} & \textbf{Function Name} & \textbf{Txns} & \textbf{Transfers} \\
    \hline
    \multirow{4}{*}{\textbf{User Cycle}}
    & \multirow{2}{*}{Yearn} & \texttt{deposit} & \num{577} & \num{2162} \\ \vspace{0.5em}
    & & \texttt{redeem} & \num{488} & \num{2127} \\
    & \multirow{2}{*}{Cian} & \texttt{optionalDeposit} & \num{159} & \num{430} \\ \vspace{0.5em}
    & & \texttt{optionalRedeem} & \num{31} & \num{166} \\
    \hline
    \multirow{3}{2.5cm}{\textbf{Strategy Management}} \vspace{0.5em}
    & Yearn & \texttt{update\_debt} & \num{707} & \num{1561} \\
    & \multirow{2}{*}{Cian} & \texttt{transferToStrategy} & \num{12} & \num{44} \\
    & & \texttt{multicall} & \num{362} & \num{1078} \\
    \hline
    \end{tabular*}
    \vspace{0.25em}
    \caption{
    Vault operations overview: Transactions (txns) and token transfers across user and strategy management cycles for each protocol.
    }
    \label{tab:core_operations_combined}
\end{table}

\section{Analysis}\label{sec:analysis}

To uncover the \emph{protocol dependencies that underpin yield aggregators' profit optimisation strategies},
we separate YA operation into two cycles, a user investment cycle and a strategy management cycle, as defined in \cref{subsec:ya-arch}.
The strategy management cycle carries the core answer to our research question:
by tracing asset flows into DeFi protocols, we identify the services involved in yield generation, fund reallocation, and yield optimisation.
The user investment cycle complements this view by characterising the use of deposited capital, the concentration of deposits across users, and the alignment of user inflows with the aggregator's reinvestment activity.

Yearn Finance and Cian contrast each other, as they employ distinct investment strategies and yield optimisation mechanisms.
For comparison of their scale, we start by reporting the statistical characteristics of deposits/withdrawals and investments.
Table~\ref{tab:deposit-withdrawal-stats} reveals significant differences in deposit patterns between the two protocols.
Cian exhibits substantially larger individual deposits with a mean (\$\CianDepositMean/) that exceeds the median (\$\CianDepositMedian/) by approximately \num{113} times, indicating a highly skewed distribution dominated by large depositors. While Yearn deposits also indicate a right skewed distribution with the mean (\$\YearnDepositMean/) exceeding the median (\$\YearnDepositMedian/) \num{48} times, it consists of more frequent but smaller individual deposits. 
While both protocols process deposits and withdrawals exceeding one million US dollars,
Cian's maximum values are notably higher, with individual deposits reaching \$\CianMaxDeposit/ compared to Yearn's \$\YearnMaxDeposit/ maximum.

\begin{table}[!htbp]
    \centering
    \small
    \begin{tabular*}{\textwidth}{@{\extracolsep{\fill}}lrrrrr}
    \hline
    \textbf{Protocol} & \textbf{Interaction} & \textbf{Median} & \textbf{Mean} & \textbf{Max} & \textbf{Total} \\

     & & & & &
    \\
    \hline
    \textbf{Yearn} & Deposits & \num{0.5}k & \num{24}k & \num{1062}k & \num{15663}k \\
    & Withdrawals & \num{3}k & \num{23}k & \num{1023}k & \num{10905}k \\
    & Investments & & & & \num{13624}k \\
    \hline
    \textbf{Cian} & Deposits & \num{3}k & \num{340}k & \num{10055}k & \num{53997}k \\
    & Withdrawals &\num{40}k & \num{565}k & \num{8599}k & \num{17515}k \\
    & Investments & & & & \num{53883}k \\
    \hline
    \end{tabular*}
    \vspace{0.25em}
    \caption{Transaction Statistics: Median, mean, and maximum values for deposits, withdrawals, and investments.}
    \label{tab:deposit-withdrawal-stats}
\end{table}

In the following, we analyse each yield aggregator separately to characterise its usage and fund movements within and across other protocols.

\subsection{Yearn Finance}\label{subsec:yearn}

We investigate network flows to and from Yearn's USDC vault to understand both its user investment patterns and internal strategy mechanics.

\subsubsection{User investment cycle.}

When users deposit USDC to Yearn, they receive yvUSDC share tokens proportional to the current share price.
As the vault generates yield, the USDC-yvUSDC exchange rate increases, enabling users to redeem more USDC than originally deposited.
Upon withdrawal, users' shares are burned and they receive the corresponding USDC amount.
If the vault has insufficient liquidity, it automatically withdraws funds from active strategies to fulfill a user's request.
We derive an annual yield of \YearnYield/, measured as the relative change in the USDC-yvUSDC exchange rate over our one-year observation window, also see Appendix~\ref{sec:appendix}.

\begin{figure}[!htb]
    \centering
    \includegraphics[width=0.9\textwidth]{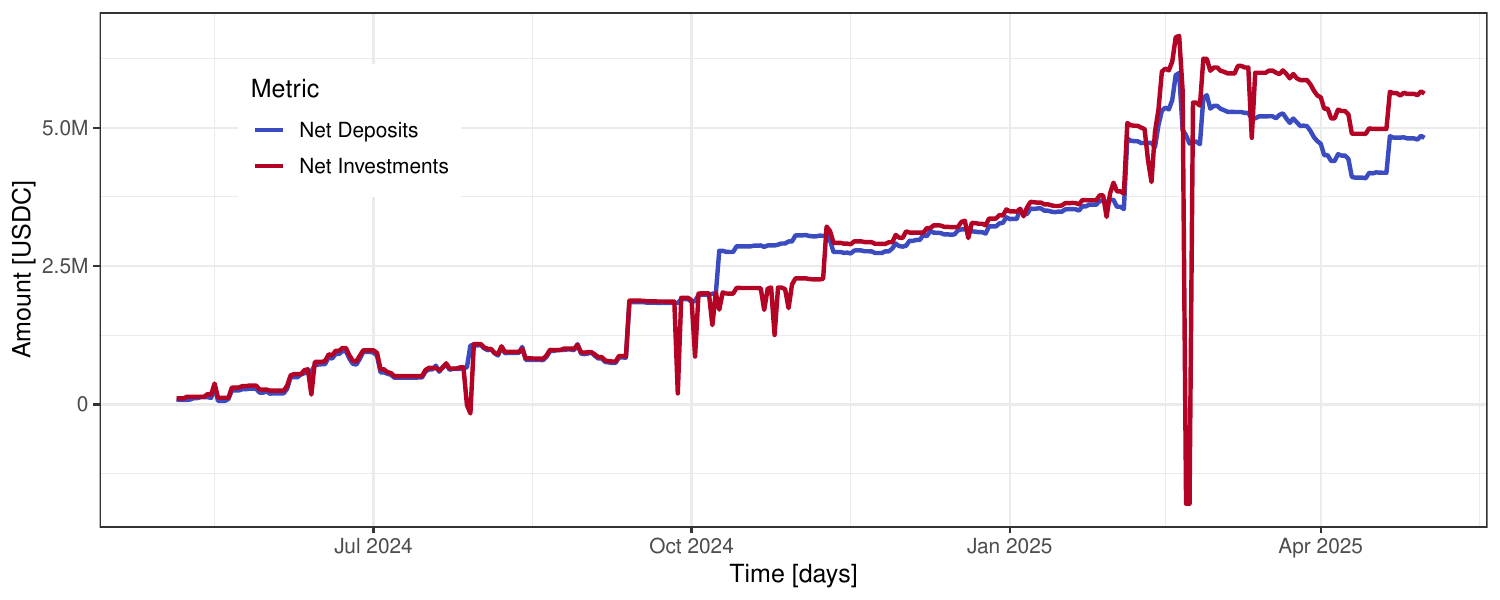}
    \caption{Yearn's net deposits and investments: Cumulative net user deposits (deposits minus withdrawals) and vault investments to external protocols (outflows minus inflows) over time.}
    \label{fig:yearn-net-deposits}
\end{figure}

We calculate net flows into the vault by subtracting withdrawals from deposits.
Figure~\ref{fig:yearn-net-deposits} illustrates the temporal evolution of both \textit{net deposits} and \textit{net investments}.
We observe steady growth in cumulative net deposits from users, indicating consistent vault expansion.
The first transaction on the Yearn protocol occurred on March 12, 2024, slightly before our observation period.

\subsubsection{Strategy management cycle.}

The Yearn vault allocates pooled capital to strategy vaults through \texttt{update\_debt} transactions.
We capture Yearn's investments to other protocols in Figure~\ref{fig:yearn-net-deposits} by calculating the net flow from Yearn's USDC vault, i.e., the outflows minus inflows in relation to other protocols.
We observe a steady increase in investments throughout the strategy management cycle, which exhibits a similar pattern to net deposits.
However, there are two inconsistencies between net deposits and investments.
First, a brief downward spike in investments in the first quarter of 2025 is quickly reversed, though the underlying reason for this reversal remains unclear.
Second, a divergence between the curves emerges in late 2024 and persists thereafter, resulting in a consistent gap beginning around the time of the downward spike.
This gap may indicate investment mechanisms not captured by our framework, such as infrequently executed capital allocation channels.
In total, \YearDepositToInvestRate/\% of deposited funds have been invested during the management cycle.

\begin{figure}[!htp]
    \centering
    \resizebox{0.9\textwidth}{!}{
    \begin{tikzpicture}[
    node distance=2cm and 1.5cm,
    vault/.style={circle, fill=white, draw=black, minimum size=1.0cm},
    contract/.style={circle, fill=white, draw=black, minimum size=0.9cm},
    edge very thin/.style={->, >=stealth, color={rgb,255:red,180;green,4;blue,38}, line width=0.4pt},
    edge thin/.style={->, >=stealth, color={rgb,255:red,180;green,4;blue,38}, line width=1.2pt},
    edge thick/.style={->, >=stealth, color={rgb,255:red,180;green,4;blue,38}, line width=2.5pt},
    edge very thick/.style={->, >=stealth, color={rgb,255:red,180;green,4;blue,38}, line width=4.0pt},
    weight/.style={font=\normalsize, sloped, above, text=black},
    label distance=2mm,
    every label/.style={font=\normalsize\bfseries, text=black, fill=white, inner sep=1pt}
]

\node[vault] (vault) at (0,0) {};
\node[contract] (compound1) at (-10,-4) {};
\node[contract] (ysusdc1) at (-8,-4) {};
\node[contract] (ysusdc2) at (-6,-4) {};
\node[contract] (aave2) at (-4,-4) {};
\node[contract] (morpho1) at (-2,-4) {};
\node[contract] (morpho2) at (0,-4) {};
\node[contract] (aave1) at (2,-4) {};
\node[contract] (aave3) at (4,-4) {};
\node[contract] (spark) at (6,-4) {};
\node[contract] (dai) at (8,-4) {};
\node[contract] (yvusdc) at (10,-4) {};
\node[contract] (compusdc) at (-10,-8) {};
\node[contract] (sky1) at (-7,-8) {};
\node[contract] (aavenode2) at (-4,-8) {};
\node[contract] (metamorph1) at (-2,-8) {};
\node[contract] (metamorph2) at (0,-8) {};
\node[contract] (aavenode1) at (2,-8) {};
\node[contract] (aavetoken) at (4,-8) {};
\node[contract] (sparkusdc) at (6,-8) {};
\node[contract] (skyusdc) at (8,-8) {};
\node[contract] (morpho) at (-1,-12) {};

\draw[edge very thick] (vault) -- node[weight, xshift=-10em] {\num{46.4}~M} (compound1);
\draw[edge thick] (vault) -- node[weight, xshift=-8em] {\num{32.0}~M} (ysusdc1);
\draw[edge thin] (vault) -- node[weight, xshift=-6em] {\num{8.4}~M} (ysusdc2);
\draw[edge thick] (vault) -- node[weight, xshift=-4em] {\num{27.6}~M} (aave2);
\draw[edge thin] (vault) -- node[weight, xshift=-2em] {\num{18.2}~M} (morpho1);
\draw[edge thin] (vault) -- node[weight] {\num{17.1}~M} (morpho2);
\draw[edge thin] (vault) -- node[weight, xshift=2em] {\num{14.4}~M} (aave1);
\draw[edge very thin] (vault) -- node[weight, xshift=4em] {\num{3.4}~M} (aave3);
\draw[edge very thin] (vault) -- node[weight, xshift=6em] {\num{1.9}~M} (spark);
\draw[edge very thin] (vault) -- node[weight, xshift=8em] {\num{1.8}~M} (dai);
\draw[edge very thin] (vault) -- node[weight, xshift=10em] {\num{0.11}~M} (yvusdc);

\draw[edge very thick] (compound1) -- node[weight, xshift=1em] {\num{46.4}~M} (compusdc);
\draw[edge thick] (ysusdc1) -- node[weight, xshift=1em] {\num{32.0}~M} (sky1);
\draw[edge thin] (ysusdc2) -- node[weight, xshift=-1em, rotate=180] {\num{8.4}~M} (sky1);
\draw[edge thick] (aave2) -- node[weight, xshift=1em] {\num{27.6}~M} (aavenode2);
\draw[edge thin] (morpho1) -- node[weight, xshift=1em] {\num{18.2}~M} (metamorph1);
\draw[edge thin] (morpho2) -- node[weight, xshift=1em] {\num{17.1}~M} (metamorph2);
\draw[edge thin] (aave1) -- node[weight, xshift=1em] {\num{14.4}~M} (aavenode1);
\draw[edge very thin] (aave3) -- node[weight, xshift=1em] {\num{3.4}~M} (aavetoken);
\draw[edge very thin] (spark) -- node[weight, xshift=1em] {\num{1.9}~M} (sparkusdc);
\draw[edge very thin] (dai) -- node[weight, xshift=1em] {\num{1.8}~M} (skyusdc);

\draw[edge thin] (metamorph1) -- node[weight, xshift=1em] {\num{18.2}~M} (morpho);
\draw[edge thin] (metamorph2) -- node[weight, xshift=-1em, rotate=180] {\num{17.1}~M} (morpho);

\node[vault, fill=blue!30, label={[align=center]below:Yearn\\Vault}] at (vault.center) {};

\node[contract, label={[align=center]below:Compound\\V3Lender}] at (compound1.center) {};
\node[contract, label={[align=center]below:Yearn\\ysUSDC}] at (ysusdc1.center) {};
\node[contract, label={[align=center]below:Yearn\\ysUSDC}] at (ysusdc2.center) {};
\node[contract, label={[align=center]below:Aave V3\\Lender}] at (aave2.center) {};
\node[contract, label={[align=center]below:Morpho\\Comp.}] at (morpho1.center) {};
\node[contract, label={[align=center]below:Morpho\\Comp.}] at (morpho2.center) {};
\node[contract, label={[align=center]below:AaveV3\\Lender}] at (aave1.center) {};
\node[contract, label={[align=center]below:AaveV3\\Lender}] at (aave3.center) {};
\node[contract, label={[align=center]below:Spark\\Lender}] at (spark.center) {};
\node[contract, label={[align=center]below:DAI\\Strategy}] at (dai.center) {};
\node[contract, label={[align=center]below:yvUSDC\\token}] at (yvusdc.center) {};

\node[contract, label={[align=center]below:Compound\\USDC\\Token}] at (compusdc.center) {};
\node[contract, label=below:Sky] at (sky1.center) {};
\node[contract, label=below:Aave] at (aavenode2.center) {};
\node[contract, label={[align=center]below:Meta\\Morpho}] at (metamorph1.center) {};
\node[contract, label={[align=center]below:Meta\\Morpho}] at (metamorph2.center) {};
\node[contract, label=below:Aave] at (aavenode1.center) {};
\node[contract, label={[align=center]below:Aave\\token\\instance}] at (aavetoken.center) {};
\node[contract, label={[align=center]below:Spark\\USDC}] at (sparkusdc.center) {};
\node[contract, label=below:SkyUSDC] at (skyusdc.center) {};

\node[contract, label=below:Morpho] at (morpho.center) {};

\end{tikzpicture}%
    }
    \caption{Yearn token transfer network:
        Total USDC inflows from Yearn to invested protocols (outflows such as rebalancing not captured) during management strategies.
        Edge weights represent amounts in millions (M) of US dollars.}
    \label{fig:yearn-update-debt-network}
\end{figure}
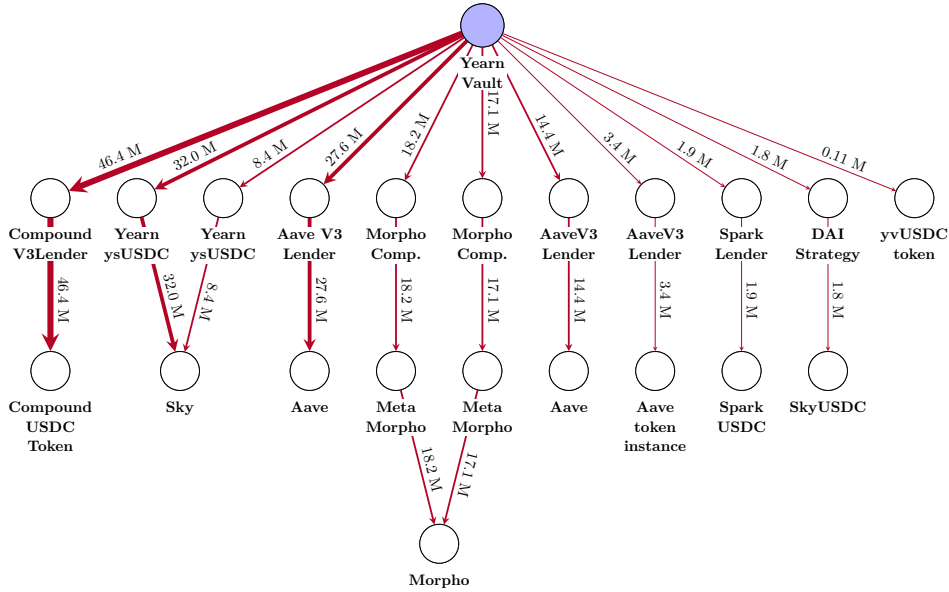

We now investigate internal investments to identify the protocol dependencies that underpin Yearn's yield generation.
When executing the \texttt{update\_debt} strategies, funds are distributed for liquidity provision, primarily to Aave, Compound, Sky, and Morpho.
The USDC token transfer network is illustrated in Figure~\ref{fig:yearn-update-debt-network},
by aggregating the flows from the Yearn vault to individual accounts that belong to yield-generating protocols.
We have labelled the accounts with tags from the Etherscan platform to provide an informed picture.
Note that these aggregated flows capture only the direction from the vault to accounts.
Therefore, when Yearn redistributes funds between protocols, that representation will contain~double~counts.

\paragraph*{Insights: protocol-specific temporal flows.}

\begin{figure}[h]
    \centering
    \includegraphics[width=0.9\textwidth]{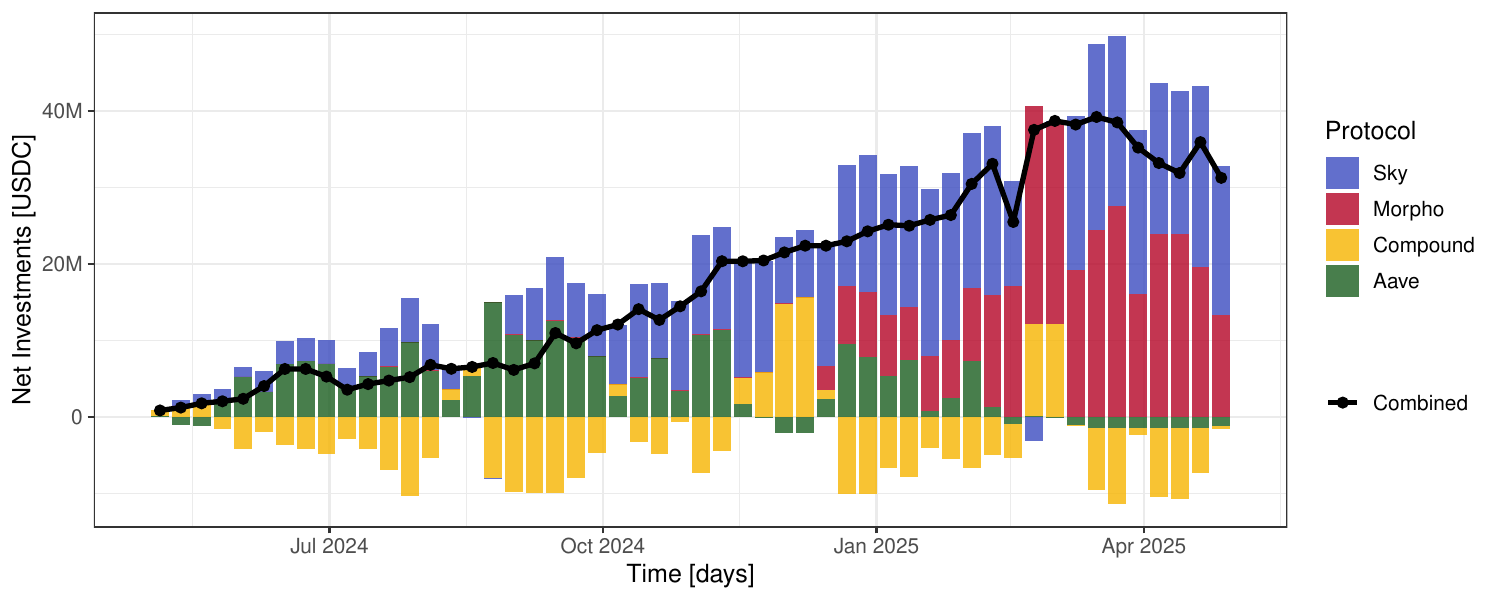}
    \caption{Protocol-specific debt flows in Yearn: Temporal evolution of net capital allocation across DeFi protocols (Aave, Compound, Sky, Morpho), showing dynamic rebalancing through \texttt{update\_debt} transactions to pursue optimal yields.}
    \label{fig:yearn-protocols-investments}
\end{figure}

We complete the previous network picture with the temporal protocol-specific flows to individual protocols in Figure~\ref{fig:yearn-protocols-investments}.
The plot reveals continuous capital rebalancing across protocols to pursue the highest returns.
Yearn also responds to the emergence of new opportunities, as evidenced by the integration of Morpho.
The targeted yield generators are mainly lending protocols, where liquidity is provided to earn interest and may receive governance token rewards.
Since Yearn existed before our observation period, the observed outflows from Compound may reflect the withdrawal of funds invested prior to our analysis window.

\paragraph*{Take-away.}
Yearn's strategy demonstrates active yield optimisation through liquidity provision and continuous rebalancing across established DeFi protocols (Aave, Compound, Sky, Morpho).
This approach results in stable annual returns of \YearnYield/ with \YearDepositToInvestRate/\% capital deployment for the USDC vault, suggesting a risk-mitigation strategy that balances yield maximisation with exposure management across multiple platforms.

\subsection{Cian}\label{subsec:cian}

We examine the staked-ETH (stETH) vault for Cian, as Cian does not offer a USDC vault within our observation period.
Analogous to Yearn, we investigate network flows to and from the vault to understand both Cian's user investment patterns and its internal strategy mechanics, which span liquid staking, restaking, and flashloan-based leverage.

\subsubsection{User investment cycle.}

Cian allows users to deposit ETH, WETH, stETH, or wstETH into a single vault.
In return, they receive ylstETH share tokens representing their proportional ownership.
The exchange rate for share tokens is documented in the protocol's event logs.
For Cian's ETH-related vault, we measure an annual yield rate of \CianYield/, extrapolated over the entire year and excluding anomalous jumps at the end of our observation period, also see Appendix~\ref{sec:appendix}.

Unlike Yearn's single-transaction withdrawals, Cian implements a two-phase redemption process:
users first submit a \texttt{requestRedeem} transaction that transfers their shares to a \textit{redeem-operator} contract,
then wait several days for their request to be processed.
The redeem-operator distributes stETH to multiple users in batched \texttt{multicall} transactions, a pattern Cian also uses to combine internal management operations (e.g., leverage execution, exchange price updates, loan repayments) into a single gas-optimised transaction.

\begin{figure}[!htb]
    \centering
    \includegraphics[width=0.9\textwidth]{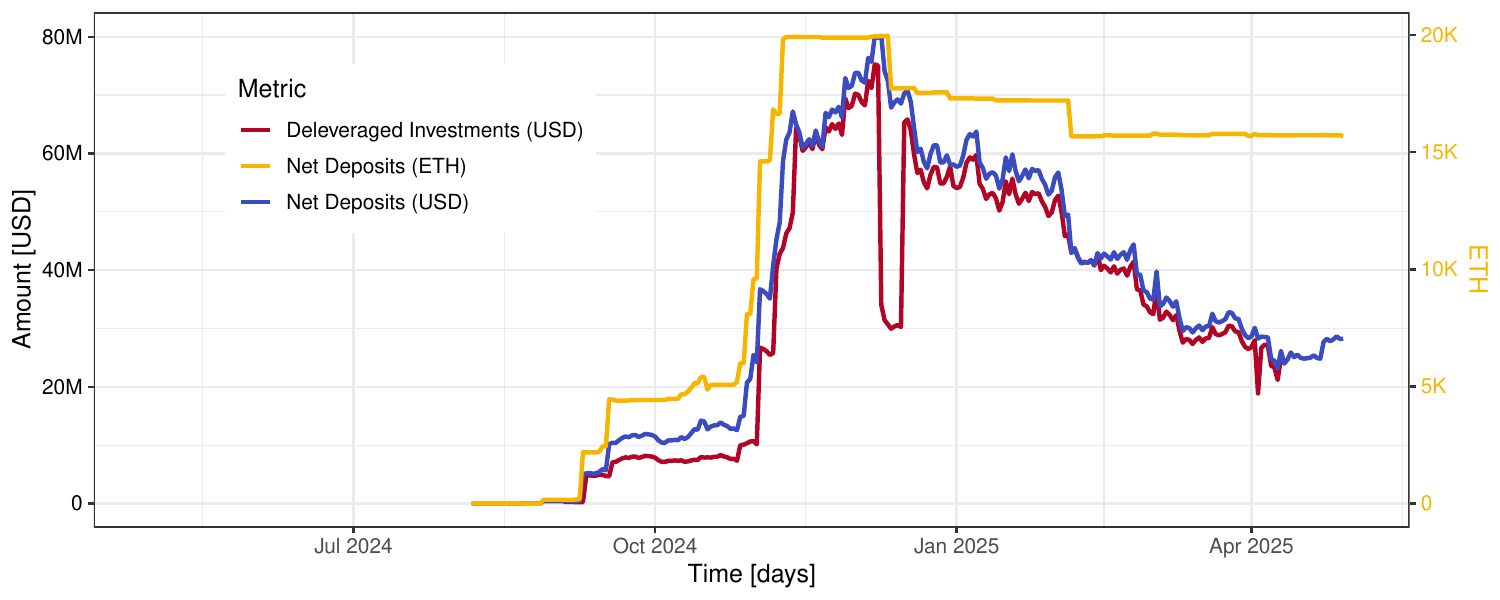}
    \caption{Cian's net deposits and deleveraged investments: Cumulative net user deposits and vault deleveraged investments to external protocols (outflows minus inflows in \texttt{transferToStrategy} transactions, excluding leverage amplification).}
    \label{fig:cian-net-deposits-investments}
    \vspace{-1em}
\end{figure}

Figure~\ref{fig:cian-net-deposits-investments} documents the cumulative net position of user deposits and investments in Cian's protocol.
Deposit values are reported in ETH, covering ETH, WETH, and stETH tokens under the assumption of value equality, and noting that wstETH deposits are automatically converted to stETH by the vault.
Both investments and deposits are converted into USD equivalents for comparative analysis with the previous YA. The USD view also illustrates how ETH/USD exchange-rate movements introduce discrepancies that can influence the realised yield of ETH-denominated strategies.
Since we measure only the inflows from the Cian vault into the strategy contracts, these investments capture the \emph{deleveraged} position, excluding capital amplified through leverage mechanisms.
Compared to Yearn, the Cian protocol was deployed later, in August 2024.
However, we find a sharp increase in deposited value in late 2024,
occurring in several large increments, which is consistent with observations from the descriptive statistics.
Thereafter the net ETH deposits remain rather constant with only occasional outflows.

\subsubsection{Strategy management cycle.}

All ETH-related assets deposited in the Cian vault are consolidated into Lido's stETH token, an ETH derivative that captures both staking positions and rewards~\cite{lido}.
The stETH then flows to strategy vaults through \texttt{transferToStrategy} transactions.
Figure~\ref{fig:cian-net-deposits-investments} also shows the temporal evolution of deleveraged investments, where we distinguish between seed capital from actual user deposits and amplified assets generated through leverage (explained in detail below).
We observe a consistent flow of deposits and corresponding investments, similar to Yearn, but the deposit curve remains constantly above the investments.
With \CianDepositToInvestRate/\% of deposited funds invested, Cian achieves an even higher investment rate than Yearn.

\begin{figure}[!t]
    \centering
    \resizebox{0.8\textwidth}{!}{
    \begin{tikzpicture}[
    node distance=2cm and 1.5cm,
    vault/.style={circle, fill=white, draw=black, minimum size=1.0cm},
    contract/.style={circle, fill=white, draw=black, minimum size=0.9cm},
    edge very thin/.style={->, >=stealth, color={rgb,255:red,180;green,4;blue,38}, line width=0.4pt},
    edge thin/.style={->, >=stealth, color={rgb,255:red,180;green,4;blue,38}, line width=1.2pt},
    edge thick/.style={->, >=stealth, color={rgb,255:red,180;green,4;blue,38}, line width=2.5pt},
    edge very thick/.style={->, >=stealth, color={rgb,255:red,180;green,4;blue,38}, line width=4.0pt},
    weight/.style={font=\footnotesize, sloped, above, text=black},
    label distance=2mm,
    every label/.style={font=\footnotesize\bfseries, text=black, fill=white, inner sep=1pt}
]

\node[vault] (vault) at (-5,0) {};
\node[contract] (strategy1) at (-5,-3) {};
\node[contract] (strategy2) at (-10,-3) {};
\node[contract] (strategy3) at (-0,-3) {};
\node[contract] (renzo) at (-5,-6) {};
\node[contract] (deposit) at (-10,-9) {};
\node[contract] (renzo1) at (-8,-9) {};
\node[contract] (renzo2) at (-5,-9) {};
\node[contract] (renzo3) at (-2,-9) {};
\node[contract] (withdraw) at (-10,-12) {};
\node[contract] (eigen) at (-5,-12) {};

\draw[edge very thick] (vault) -- node[weight, xshift = 0.5em] {\num{46.8}~M} (strategy1);
\draw[edge very thin] (vault) -- node[weight, xshift = -3em] {0.0001~M} (strategy2);
\draw[edge thin] (vault) -- node[weight, xshift = 3em] {\num{6.8}~M} (strategy3);

\draw[edge very thick] (strategy1) -- node[weight, xshift = 0em] {\num{46.8}~M} (renzo);

\draw[edge very thin] (renzo) -- node[weight, xshift = -4em] {\num{1.7}~M} (deposit);
\draw[edge thick] (renzo) -- node[weight, xshift = -2em] {\num{20.7}~M} (renzo1);
\draw[edge thin] (renzo) -- node[weight, xshift = 0em] {\num{8.7}~M} (renzo2);
\draw[edge thick] (renzo) -- node[weight, xshift = 2em] {\num{15.6}~M} (renzo3);

\draw[edge very thin] (deposit) -- node[weight, xshift = 0em] {\num{1.7}~M} (withdraw);
\draw[edge thick] (renzo1) -- node[weight, xshift = 0em] {\num{20.7}~M} (eigen);
\draw[edge thin] (renzo2) -- node[weight, xshift = 0em] {\num{8.7}~M} (eigen);
\draw[edge thick] (renzo3) -- node[weight, xshift = -0em] {\num{15.6}~M} (eigen);

\node[vault, fill=blue!30, label={[align=center]left:Cian\\Vault}] at (vault.center) {};

\node[contract, label={[align=center]left:Strategy\\AAVEV3\\EzETH}] at (strategy1.center) {};
\node[contract, label={[align=center]left:Strategy\\Mellow\\Steak.}] at (strategy2.center) {};
\node[contract, label={[align=center]right:Strategy\\AAVE V3\\LIDO}] at (strategy3.center) {};

\node[contract, label={[align=center]left:Renzo\\Protocol}] at (renzo.center) {};

\node[contract, label={[align=center]left:Deposit\\Queue}] at (deposit.center) {};
\node[contract, label={[align=center]right:Renzo\\Protocol}] at (renzo1.center) {};
\node[contract, label={[align=center]right:Renzo\\Protocol}] at (renzo2.center) {};
\node[contract, label={[align=center]right:Renzo\\Protocol}] at (renzo3.center) {};

\node[contract, label={[align=center]left:Renzo\\Withdraw\\Queue}] at (withdraw.center) {};
\node[contract, label={[align=center]right:EigenLayerStrategies}] at (eigen.center) {};

\end{tikzpicture}%
}
\caption{Cian token transfer network showing stETH transfers to strategies. Total
inflows from Cian to invested protocols (outflows such as rebalancing not cap-
tured). Edge weights represent transfer amounts in millions of US dollars (M).}
\label{fig:cian-transfer-to-strategy}
\end{figure}
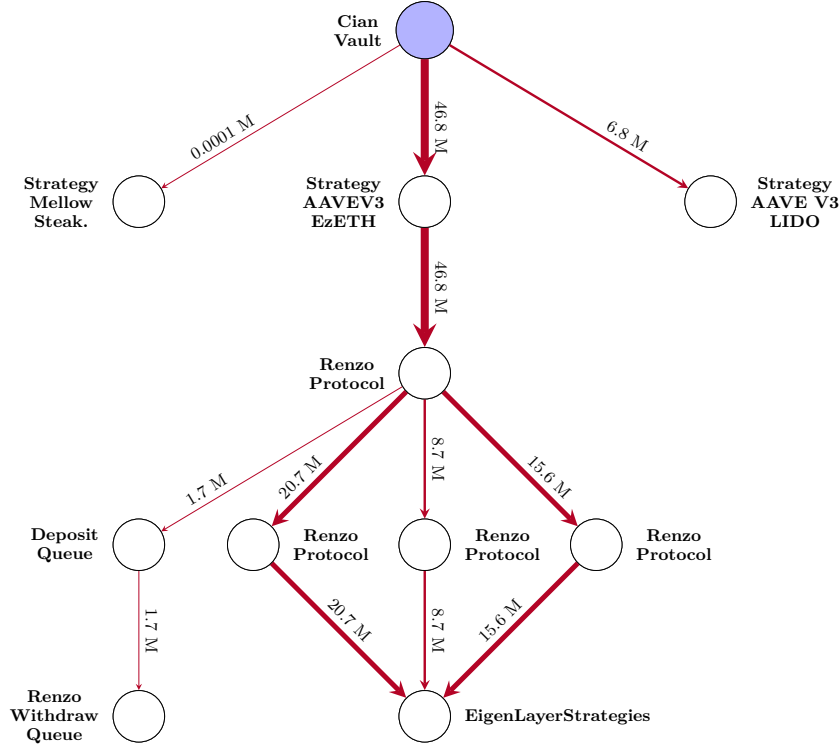

Figure~\ref{fig:cian-transfer-to-strategy} illustrates the transfer network for the stETH token when executing \\ \texttt{transferToStrategy} transactions.
The stETH tokens are first deposited into the liquid restaking protocol Renzo, and the underlying stake is subsequently restaked through EigenLayer.
This routing potentially stacks additional reward layers on top of stETH's underlying validator rewards from Lido: Renzo's LRT issuance and restaking rewards from EigenLayer.

\paragraph*{Insights: Cian's leverage strategy.}

Beyond multi-protocol routing, Cian employs a sophisticated flashloan-enabled instance of looping.
This mechanism amplifies user deposits into leveraged positions, maximising staking rewards while introducing additional risk.

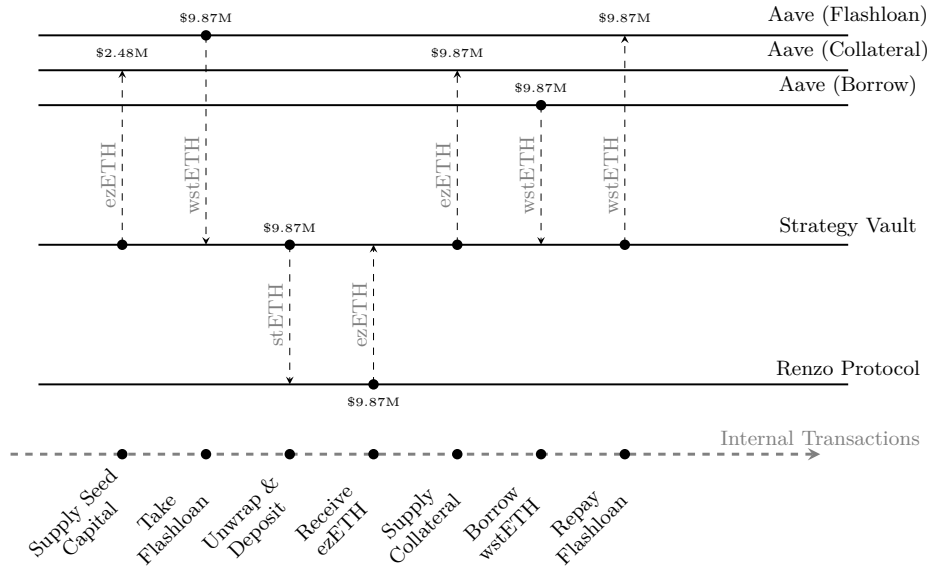
\begin{figure}[!htb]
    \centering
    \resizebox{0.9\textwidth}{!}{
    \begin{tikzpicture}[xscale=0.8,>=stealth]
    \tikzstyle{timepoint}=[circle, fill=black, inner sep=1.5pt]
    \tikzstyle{timeline}=[thick, ->]

    \draw[thick] (-3.5,7) -- (11,7) node[above, font=\footnotesize] {Aave (Flashloan)};
    \draw[thick] (-3.5,6.5) -- (11,6.5) node[above, font=\footnotesize] {Aave (Collateral)};
    \draw[thick] (-3.5,6) -- (11,6) node[above, font=\footnotesize] {Aave (Borrow)};
    \draw[thick] (-3.5,4) -- (11,4) node[above, font=\footnotesize] {Strategy Vault};
    \draw[thick] (-3.5,2) -- (11,2) node[above, font=\footnotesize] {Renzo Protocol};

    \draw[very thick, ->, gray, dashed] (-4,1) -- (10.5,1) node[above, font=\footnotesize] {Internal Transactions};

    \node[timepoint] at (-0.5,7) {};     
    \node[timepoint] at (5.5,6) {};     

    \node[timepoint] at (-2,4) {};    
    \node[timepoint] at (1,4) {};     
    \node[timepoint] at (4,4) {};     
    \node[timepoint] at (7,4) {};    

    \node[timepoint] at (2.5,2) {};     

    \node[timepoint] at (-2,1) {};
    \node[timepoint] at (-0.5,1) {};
    \node[timepoint] at (1,1) {};
    \node[timepoint] at (2.5,1) {};
    \node[timepoint] at (4,1) {};
    \node[timepoint] at (5.5,1) {};
    \node[timepoint] at (7,1) {};

    \draw[->, dashed] (-2,4) -- (-2,6.5);
    \node[font=\footnotesize, text=gray, rotate=90] at (-2.25,5) {ezETH};
    \node[font=\tiny, text=black] at (-2,6.75) {\$2.48M};

    \draw[->, dashed] (-0.5,7) -- (-0.5,4);
    \node[font=\footnotesize, text=gray, rotate=90] at (-0.75,5) {wstETH};
    \node[font=\tiny, text=black] at (-0.5,7.25) {\$9.87M};

    \draw[->, dashed] (1,4) -- (1,2);
    \node[font=\footnotesize, text=gray, rotate=90] at (0.75,3) {stETH};
    \node[font=\tiny, text=black] at (1,4.25) {\$9.87M};

    \draw[->, dashed] (2.5,2) -- (2.5,4);
    \node[font=\footnotesize, text=gray, rotate=90] at (2.25,3) {ezETH};
    \node[font=\tiny, text=black] at (2.5,1.75) {\$9.87M};

    \draw[->, dashed] (4,4) -- (4,6.5);
    \node[font=\footnotesize, text=gray, rotate=90] at (3.75,5.) {ezETH};
    \node[font=\tiny, text=black] at (4,6.75) {\$9.87M};

    \draw[->, dashed] (5.5,6) -- (5.5,4);
    \node[font=\footnotesize, text=gray, rotate=90] at (5.25,5) {wstETH};
    \node[font=\tiny, text=black] at (5.5,6.25) {\$9.87M};

    \draw[->, dashed] (7,4) -- (7,7);
    \node[font=\footnotesize, text=gray, rotate=90] at (6.75,5) {wstETH};
    \node[font=\tiny, text=black] at (7,7.25) {\$9.87M};

    \node[font=\footnotesize, below, align=center, rotate=45, xshift = -25, yshift = 10] at (-2,0.7) {Supply Seed\\Capital};
    \node[font=\footnotesize, below, align=center, rotate=45, xshift = -25, yshift = 10] at (-0.5,0.7) {Take\\Flashloan};
    \node[font=\footnotesize, below, align=center, rotate=45, xshift = -25, yshift = 10] at (1,0.7) {Unwrap \&\\Deposit};
    \node[font=\footnotesize, below, align=center, rotate=45, xshift = -25, yshift = 10] at (2.5,0.7) {Receive\\ezETH};
    \node[font=\footnotesize, below, align=center, rotate=45, xshift = -25, yshift = 10] at (4,0.7) {Supply\\Collateral};
    \node[font=\footnotesize, below, align=center, rotate=45, xshift = -25, yshift = 10] at (5.5,0.7) {Borrow\\wstETH};
    \node[font=\footnotesize, below, align=center, rotate=45, xshift = -25, yshift = 10] at (7,0.7) {Repay\\Flashloan};

\end{tikzpicture}
    }
    \caption{Flashloan-enabled leveraged staking: A single transaction deposits seed capital and flashloan-funded ezETH collateral to Aave, borrows wstETH, and repays the flashloan.}

    \label{fig:flashloan-spacetime}
\end{figure}

We illustrate this mechanism through an exemplary transaction\footnote{Hash: \texttt{0xf2ce4bc0e749465221788d32877f3bfca1163d9eed5d4ff57e8d8eaee069c649}} in Figure~\ref{fig:flashloan-spacetime}.
Starting with \$\num{2.48}M in ezETH as seed collateral on Aave, the strategy takes a flashloan of \$\num{9.87}M in wstETH that must be repaid within the same transaction.
The borrowed wstETH is unwrapped to stETH and deposited into Renzo, which returns \$\num{9.87}M in ezETH tokens.
These newly acquired ezETH tokens are supplied as additional collateral to Aave, bringing total collateral to \$\CianCollateralAfterLeverage/, approximately \CianLeverageMultiplier/$\times$ the initial seed capital.
The ezETH collateral on Aave earns staking rewards from three layers: Ethereum validator rewards via Lido, restaking rewards through Renzo, and additional yields from EigenLayer.
Against this amplified collateral, the strategy borrows \$\num{9.87}M in wstETH from Aave to repay the flashloan, closing the loop.
The final position holds \$\num{12.35}M in ezETH collateral while owing \$\num{9.87}M in wstETH debt.
The debt amount equals the additional collateral generated through the flashloan cycle.
By subtracting the debt from total collateral (i.e., $\$\num{12.35}M - \$\num{9.87}M = \$\num{2.48}M$), we recover the original seed capital, which we term \textit{deleveraged investments} in Figure~\ref{fig:cian-net-deposits-investments}.
This leveraged position constitutes the effective capital exposed to ezETH staking rewards, amplifying both potential returns and liquidation risks well beyond the original seed deposit.

\paragraph*{Summary.}
Cian's strategy combines several profit sources by routing funds through multiple protocols.
On top of this multi-protocol routing, Cian maximises staking rewards through flashloan-enabled recursive leverage ($\approx \CianLeverageMultiplier/\times$).
This sophisticated approach achieves returns of \CianYield/ for the ETH-related vault but introduces significant complexity and amplified liquidation risks.
The outcome of this composition is a layered set of protocol dependencies spanning liquid staking (Lido), liquid restaking (Renzo), restaking (EigenLayer), and lending with flashloans (Aave), further extended through wrapped derivative assets (stETH, wstETH, ezETH) that introduce additional asset-level couplings.

\section{Discussion}\label{sec:disc}

\subsection{Yield Generation}

Our analysis surfaces two markedly different answers to the research question \emph{what protocol dependencies underpin yield aggregators' profit optimisation strategies}: Cian and Yearn use distinct sets of DeFi protocol categories to generate returns, resulting in different levels of complexity and risk.
While Yearn primarily invests user deposits as collateral into lending services,
Cian follows a more sophisticated process that stakes and wraps deposited assets and additionally employs leverage mechanisms through flashloans, producing a denser and more layered set of protocol dependencies.
Yearn exhibits steady growth in user deposits throughout our observation period, while the later-introduced Cian is characterised by a sharp rise in deposited capital concentrated in several large increments.

For both protocols, the two cycles, user investment cycle and the strategy management cycle, are relatively synchronised.
This synchronization is further evidenced by high investment-to-deposit ratios of \YearDepositToInvestRate/\% for Yearn and \CianDepositToInvestRate/\% for Cian.
However, these cycles are not completely independent.
For Yearn, when the vault lacks sufficient liquidity, it automatically withdraws funds from strategies to fulfill user withdrawal requests, demonstrating how the user cycle can trigger liquidity removal from strategy investments.
Both protocols exhibit brief downward-spike interruptions that are reverted rather quickly.
However, distinct from Cian, where the deleveraged net-investment curve always remains below net deposits, this relationship reverses at some point for Yearn.
We hypothesize possible explanations, such as reinvested yield that increases investment above deposits or rare, additional investment mechanisms not captured in our analytical framework.

When comparing the interest rates achieved by the aggregators,
the more sophisticated and riskier strategy of Cian yields \CianYield/, while Yearn achieves \YearnYield/.
Despite its added complexity and leverage mechanisms, Cian's returns are lower than Yearn's during our observation period.
We analyse the USDC vault for Yearn and the stETH vault for Cian, as Cian does not offer a USDC vault in our observation period.
This comparison involves different asset types: Yearn's stablecoin vault versus Cian's native ETH derivative vault.
For ETH-based assets, exchange rate fluctuations can amplify or diminish the manifested profit when measured in US dollars, introducing additional volatility not present in stablecoin strategies. To contextualise these returns, we compare them against two baselines over the same observation period: the average Federal Funds Effective Rate of \FedFundsRate/~\cite{fedfunds}, representing a conventional risk-free benchmark, and the average APY of \STETHBaseline/ for simply holding stETH~\cite{defillama_steth_pool}, representing the passive baseline for ETH-based yield.
Yearn's \YearnYield/ exceeds both baselines, while Cian's \CianYield/ falls below the risk-free rate and only marginally surpasses simply holding stETH, offering little reward for its added complexity and risk.

\subsection{Risk Exposure of Yield Aggregators}

\begin{figure}[h]
    \centering
    \resizebox{\textwidth}{!}{
    \input{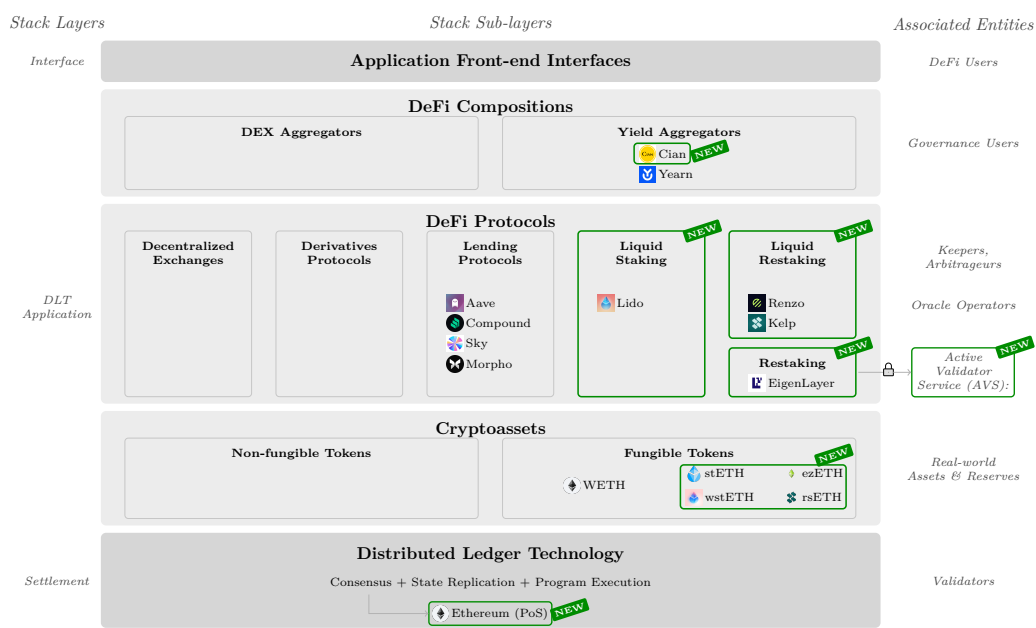}
    }
    \caption{Augmented DeFi Stack Reference Model (DSR), adapted from~\cite{Auer:2023a} and extended with the protocol types and instances relevant to this study.
        Layers from the original DSR are shown in light-gray, the augmented \emph{System Infrastructure} layer in dark-gray.
        \emph{Liquid Staking}, \emph{Restaking}, and \emph{Liquid Restaking} are introduced as new DeFi protocol types alongside DEX, Derivatives, and Lending. Icons denote concrete protocol/asset instances.}
    \label{fig:dsr_model}
\end{figure}

Yearn and Cian rely on composable, multi-protocol investment strategies that expose these yield aggregators to a broad set of interconnected risks. To systematise these exposures, we adapt the DeFi Stack Reference (DSR) Model of Auer et al.~\cite{Auer:2023a} (see Figure~\ref{fig:dsr_model}), extending it with three new DeFi protocol types relevant to our analysis, \emph{Liquid Staking}, \emph{Restaking}, and \emph{Liquid Restaking}, and populating each layer with the protocols and assets touched in our study. We additionally introduce \emph{actively validated services} (AVSs) as a new associated entity, capturing the linkage to external platforms and DLTs that restaked capital secures. Building on this architecture, we develop a layered DeFi risk propagation framework (see Table~\ref{tab:risk_taxonomy}) that maps key risk categories onto the underlying protocol stack.

Our analysis reveals complex structural dependencies among the DeFi protocols involved in yield aggregation. Therefore, we conceptualise DeFi risk as a propagation process across the stack layers, where localised shocks originating at the settlement, asset, or protocol layer may transmit through collateral dependencies, liquidity linkages, and leverage structures. Importantly, our empirical results indicate that the architecture of the aggregation strategy itself shapes the intensity and transmission pathways of these risks. Whereas Yearn primarily relies on comparatively simple lending-based strategies, Cian combines staking, wrapped assets, lending, and flashloan-enabled leverage into tightly coupled recursive investment structures spanning multiple layers of the DeFi stack.

First, yield aggregators are exposed to \textbf{technology risks}, which constitute the foundational vulnerabilities of the DeFi stack. Aggregators inherit these baseline exposures from the underlying platforms they integrate~\cite{Cousaert2022}, including smart contract exploits, oracle manipulation, validator slashing conditions, and exit delays. Such risks typically originate in the technology and protocol components of the architecture and propagate through collateral and liquidity dependencies. For example, an AVS-imposed slashing event at the protocol layer reduces the backing value of LRTs at the asset layer, weakening their effectiveness as collateral in lending protocols~\cite{alexander2024leveraged}. Although the slashed stake is removed on-chain immediately, LRT holders can realize the loss only through a multi-step withdrawal: unwrapping the LRT into its underlying LST, and then exiting the validator queue, which can span weeks. Because closing the gap via redemption requires locking capital across this multi-week exit window, arbitrage cannot quickly realign the secondary-market price with the reduced backing value, and the resulting depeg can substantially exceed the original slashing magnitude—triggering liquidations of LRT collateral and forced deleveraging in aggregation strategies.

\begin{table}[t]
\centering
\footnotesize
\begin{threeparttable}
\begin{tabular}{L{3.7cm}C{2cm}C{2cm}C{2cm}C{2cm}}
\toprule
\multirow{2}{*}{Risk} & \multicolumn{4}{c}{Stack layers} \\
\cmidrule(lr){2-5}
& DLT & Assets & Protocols & Compositions\\
\midrule
\multicolumn{5}{l}{\textbf{Technology risks}}\\
Code defects & \ding{55} & \ding{55} & \ding{55}  & \ding{55}\\
Oracle mispricing & & & \ding{55} & \\
Slashing & \ding{55} &  & \ding{55} & \\
Exit delays & \ding{55} & & \ding{55} & \ding{55}\\
\midrule
\multicolumn{5}{l}{\textbf{Market and financial risks}}\\
Price volatility & & \ding{55} &  & \\
Correlation risk & & \ding{55} &  & \\
Token depegs & & \ding{55} & \ding{55} & \\
Liquidity risk & & \ding{55} & \ding{55} & \ding{55} \\
Liquidation cascades & &  & \ding{55} & \ding{55}\\
Leverage loops & &  & \ding{55} & \ding{55}\\
\midrule
\multicolumn{5}{l}{\textbf{Structural risks}}\\
Concentration risk & \ding{55} & \ding{55} & \ding{55} & \ding{55}\\
Governance risk & \ding{55} & & \ding{55} & \ding{55}\\
\bottomrule
\end{tabular}
\caption{Risk exposure of yield aggregators. This table maps major DeFi risks to the layers where they primarily originate. Depending on their magnitude, shocks may propagate across the DeFi stack.}
\label{tab:risk_taxonomy}
\end{threeparttable}
\end{table}

Building on these foundational vulnerabilities, \textbf{market and financial risks} arise from the interplay of collateral valuation, liquidity, and leverage. A central insight from the literature~\cite{alexander2024leveraged,Gogol2024sok} is that both liquid staking tokens (LSTs) and, more prominently, liquid restaking tokens (LRTs) are exposed to depeg risk, i.\,e., deviations from the underlying asset price that can trigger liquidation when these tokens serve as collateral. Because lending protocols continuously revalue collateral via on-chain oracles and expose any undercollateralized position to permissionless liquidation by external keepers, such asset-layer shocks transmit directly into lending markets and from there into aggregation strategies that rely on recursive borrowing~\cite{alexander2024leveraged}. Empirical evidence from Compound further shows that such shocks need not stay confined to the affected asset. Liquidations triggered by a collateral price drop can leave connected pools with bad loans and propagate insolvency across the lending protocol~\cite{Tovanich2023}.

These dynamics are especially relevant for Cian’s strategy design. Although Cian’s USD performance depends on the ETH/USD exchange rate through its stETH-based vault, the more critical exposure is the peg between ETH and its staked and restaked derivatives. In the short run, this peg is held in place by arbitrage trading on decentralised exchanges, yet for many LSTs and LRTs the relevant DEX pools have comparatively low liquidity~\cite{alexander2024leveraged}, leaving the peg vulnerable to even modest sell pressure. Recursive leverage compounds the problem, because the same ETH-based collateral is repeatedly looped through staking and lending, even moderate price movements can trigger deleveraging cascades and liquidity withdrawals across the interconnected positions.

Recent market episodes illustrate these cross-layer risk propagation mechanisms. The KelpDAO/rsETH incident~\cite{KelpHack2026}, for example, originated at the infrastructure layer through vulnerabilities associated with cross-chain bridge operations. Although the initial disruption emerged in the infrastructure components of the DeFi architecture, it rapidly propagated through the stack. The compromised or invalid rsETH collateral impaired confidence in the asset layer and subsequently generated liquidity stress at the protocol layer when rsETH was used as collateral on Aave. In response, large actors initiated coordinated withdrawals and deleveraging, causing billions in outflows and pushing several lending markets toward near-full utilisation. This incident demonstrates how infrastructure-level disruptions may cascade through collateral dependencies, liquidity withdrawals, and automated liquidation mechanisms, ultimately affecting aggregation strategies. Although, we have not seen any major events on yield aggregators so far, this recent incidents shows how exploits can propagate and affect the DeFi ecosystem.

\textbf{Structural risks} emerge from the composability of DeFi architectures themselves. Concentration risk is related to the dominance of a small number of LST providers, validators, AVSs, or liquidity pools. Such concentration may transform localised shocks into broader market failures when large actors unwind positions or strategically influence prices. In extreme scenarios, these dynamics can generate contagion across multiple layers of the DeFi stack, affecting not only individual protocols but the resilience of the broader ecosystem.

Our empirical findings suggest that these amplification mechanisms are likely more pronounced in Cian than in Yearn. Cian’s sharp increases in deposited capital, concentrated in several large increments, indicate a higher degree of depositor concentration and therefore greater vulnerability to coordinated withdrawals or strategic liquidity movements by large actors. Combined with recursive leverage structures and ETH-denominated collateral exposure, such concentration may intensify liquidity spirals and increase the probability of cascade dynamics during periods of market stress. By contrast, Yearn’s more gradual deposit growth during our observation period and its simpler lending-oriented architecture imply comparatively lower structural fragility.

The more complex and risk-intensive architecture of Cian does not translate into superior realised returns during the observation period. Despite employing leverage mechanisms and recursive staking strategies, Cian underperforms Yearn’s comparatively simpler lending-based vault. This finding suggests that additional composability and leverage may increase structural fragility and systemic risk exposure without necessarily generating proportional yield advantages.

\subsection{Limitations and Future Work}
Our study faces several limitations that future work should address.
First, we examine a subset of two yield aggregators and analysed one of their available vaults over a limited time horizon, which may not fully represent the entire ecosystem.
Future work should extend this analysis to additional protocols, vault types and extend the observation windows to achieve broader coverage.
To support such extensions, we publicly release the code and data used in this study.\footnote{\url{\RepoUrl/}}
Second, while we provide detailed insights into leveraging mechanisms through representative transactions, our analysis does not exhaustively document all variations and edge cases of these complex strategies.
We believe that understanding the fundamental structure of these mechanisms is essential for assessing systemic risks and interdependencies, which remains an important direction for future research.
Third, our operational risk taxonomy does not address fraud-related exposures such as rug pulls, Ponzi-like structures, or pump-and-dump schemes that fall outside the protocols studied here but represent a distinct vector for any aggregation strategy interacting with unverified counterparties~\cite{CarpentierDesjardins2025}.

\section{Conclusion}\label{sec:concl}

This study examined how DeFi yield aggregators operate, their protocol dependencies, and capital flow patterns by analysing Yearn Finance and Cian Yield Layer over a one-year period.
We find that the two platforms embody fundamentally different approaches to yield generation.
Yearn pursues a well-established interest-bearing strategy, providing collateral and dynamically shuffling user deposits across mature lending protocols to earn interest rewards.
Cian, in contrast, exploits recent DeFi innovations by routing funds through liquid (re)staking derivatives and amplifying these positions with flashloan-based leverage.
While our empirical analysis shows that the more complex and risk-intensive architecture of Cian does not translate into superior realised returns during the observation period, it introduces substantially more interconnections across the layers of the DeFi stack.
These additional dependencies enlarge the surface for cross-layer risk propagation, making such strategies more vulnerable to shocks that can cascade through the underlying infrastructure.
Our findings highlight that strategic complexity in yield aggregation does not necessarily translate into higher returns, but it materially increases exposure to systemic risk in the composable DeFi ecosystem.

\ifdefined\IsShortPaper
\section*{Supplemental Material}\label{sec:supplement}
The full appendix (manual deposit/withdrawal transaction hashes, vault function signatures, cumulative deposit/withdrawal distributions, vault exchange-rate signals, and the deposit-vs-investment comparison) is available in the extended version accompanying this submission and at \url{\RepoUrl/}.

\appendix
\refstepcounter{section}\label{sec:appendix}
\fi

\bibliography{references}

\ifdefined\IsShortPaper\else
\appendix
\clearpage

\section{Appendix}\label{sec:appendix}

Table~\ref{tab:transaction_hashes} lists the transaction hashes of the manual deposit/withdrawal cycles we executed against each yield aggregator to identify the core vault operations described in Section~\ref{sec:data}.

\begin{table}[htbp]
    \centering
    \small
    \begin{tabular}{ll}
    \hline
    \textbf{Protocol} & \textbf{Transaction Hash} \\
    \hline
    \multirow{4}{*}{Cian}
    & \texttt{0x13a8feb5ac4eb62618ed22f0b8a5d6623ffcba35f0692bcc30d81a71e1824305}  \\
    & \texttt{0xaaeff91b776c4896021a3e490518fa781e8628999c9a3a696634df1c92e815ff} \\
    & \texttt{0x54b8bd24c062dc616eb64683d466c3d269ee27948e4fbc491e2bea69df56cff0} \\
    & \texttt{0x40e6bfbc1669c49af1404e128c4e557e56db9ec7522e5a76b09b5cd1041adb0d} \\
    \hline
    \multirow{2}{*}{Yearn}
    & \texttt{0x32b15cce70391b491e0c0ef357e2883db4a0e1eea5633427c10cebba10121e57} \\
    & \texttt{0xe47111e9bed7b7a3d76a7452fe8b713d3056962067a3e13871600357b54067e5} \\
    \hline
    \end{tabular}
    \caption{Transaction Hashes of Manual Interactions with Yield Aggregators}
    \label{tab:transaction_hashes}
\end{table}

Table~\ref{tab:function_signatures} lists the 4-byte function signatures of the core vault operations (deposit, redeem, and internal rebalancing entry points) used to filter and classify on-chain interactions with each protocol.

\begin{table}[htbp]
    \centering
    \begin{tabular*}{\textwidth}{@{\extracolsep{\fill}}llr}
    \hline
    \textbf{Protocol} & \textbf{Function Name} & \textbf{Function Signature} \\
    \hline
    \multirow{3}{*}{Yearn} & \texttt{deposit} & 0x6e553f65 \\
    & \texttt{redeem} & 0x9f40a7b3, 0xba087652 \\
    & \texttt{update\_debt} & 0xba54971f, 0x0aeebf55 \\
    \hline
    \multirow{4}{*}{Cian} & \texttt{optionalDeposit} & 0x32507a5f \\
    & \texttt{optionalRedeem} & 0xa7b73254 \\
    & \texttt{transferToStrategy} & 0xba8bfa2a \\
    & \texttt{multicall (leverage)} & 0xda5b4ffd \\
    \hline
    \end{tabular*}
    \caption{Function Signatures for Core Vault Operations}
    \label{tab:function_signatures}
\end{table}

Figure~\ref{fig:yearn-protocols-4x1} shows the cumulative probability distributions of withdrawals and deposits for Cian and Yearn. The distributions appear as long right tails. Yearn's curves are smooth and gradual, suggesting many small transactions. Cian's contain more discrete jumps, suggesting fewer, larger transactions such as whale activity.

\begin{figure}[h]
    \centering
    \includegraphics[width=0.8\textwidth]{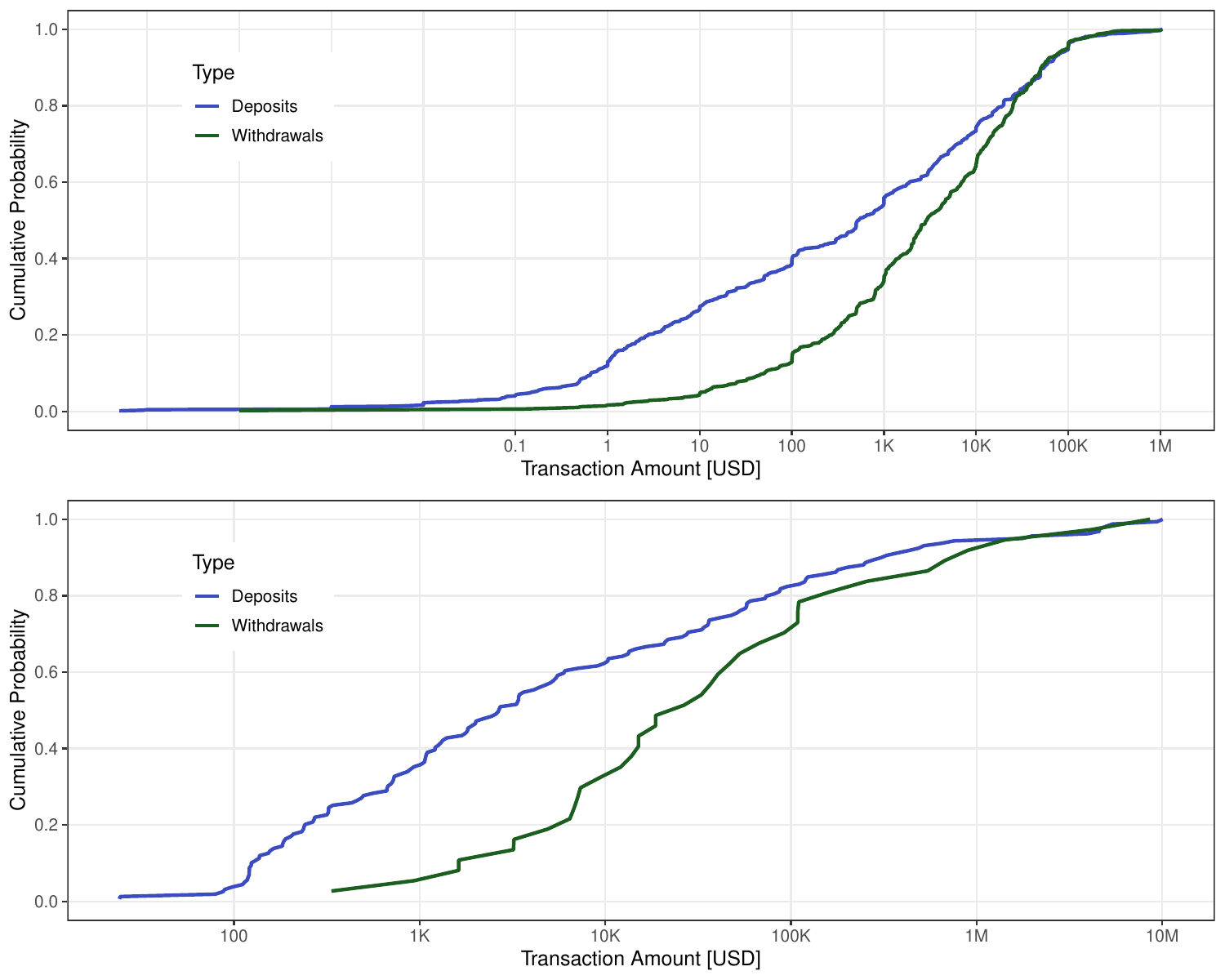}
    \caption{Cumulative distributions of vault transactions: Empirical cumulative probabilities of deposit and withdrawal sizes for Yearn (top) and Cian (bottom). }
    \label{fig:yearn-protocols-4x1}
\end{figure}

Figures~\ref{fig:yearn-exchange-rate} and~\ref{fig:cian-exchange-rate} plot the share-to-asset exchange rate of Yearn's USDC vault (USDC--yvUSDC) and Cian's ETH vault (ETH--ylstETH), the on-chain signal underlying our realised-yield computation in Section~\ref{sec:analysis}.

\begin{figure}[h]
    \centering
    \includegraphics[width=0.8\textwidth]{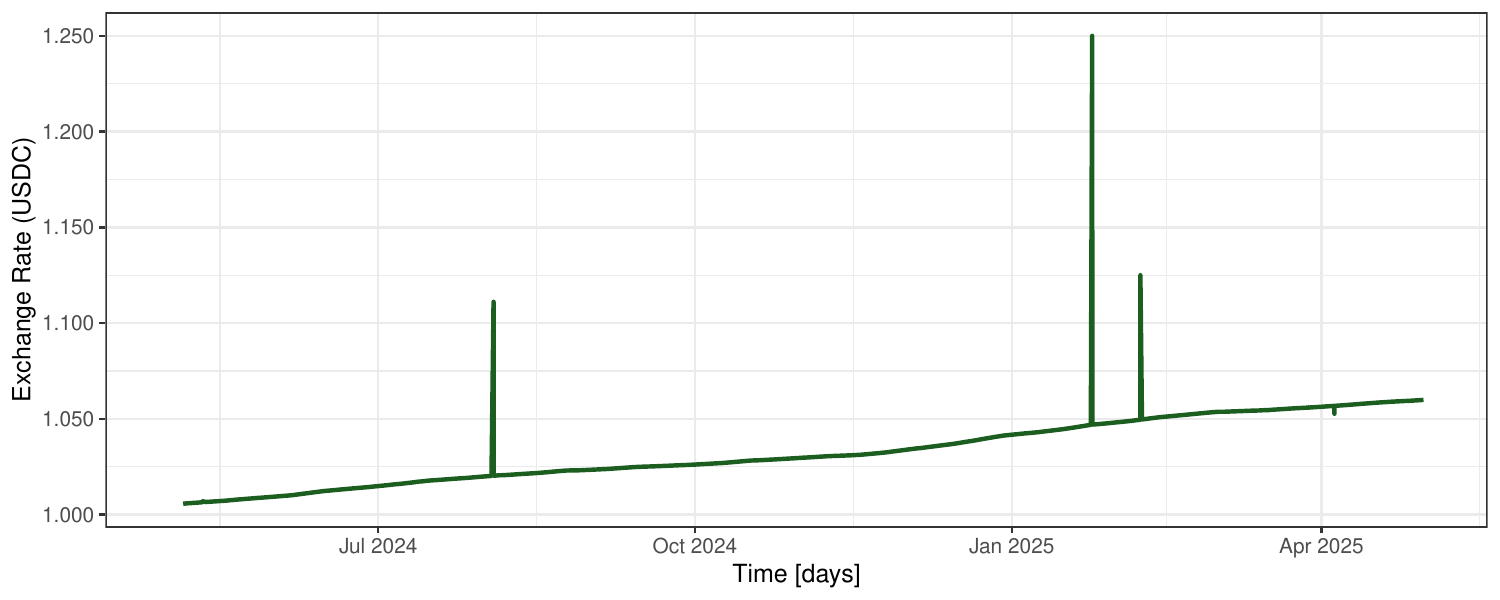}
    \caption{Yearn USDC-vault exchange rate: Daily evolution of the share-to-asset exchange rate (USDC--yvUSDC) of Yearn's USDC vault over the observation period, the on-chain signal underlying the realised-yield computation in Section~\ref{sec:analysis}.}
    \label{fig:yearn-exchange-rate}
\end{figure}

\begin{figure}[h]
    \centering
    \includegraphics[width=0.8\textwidth]{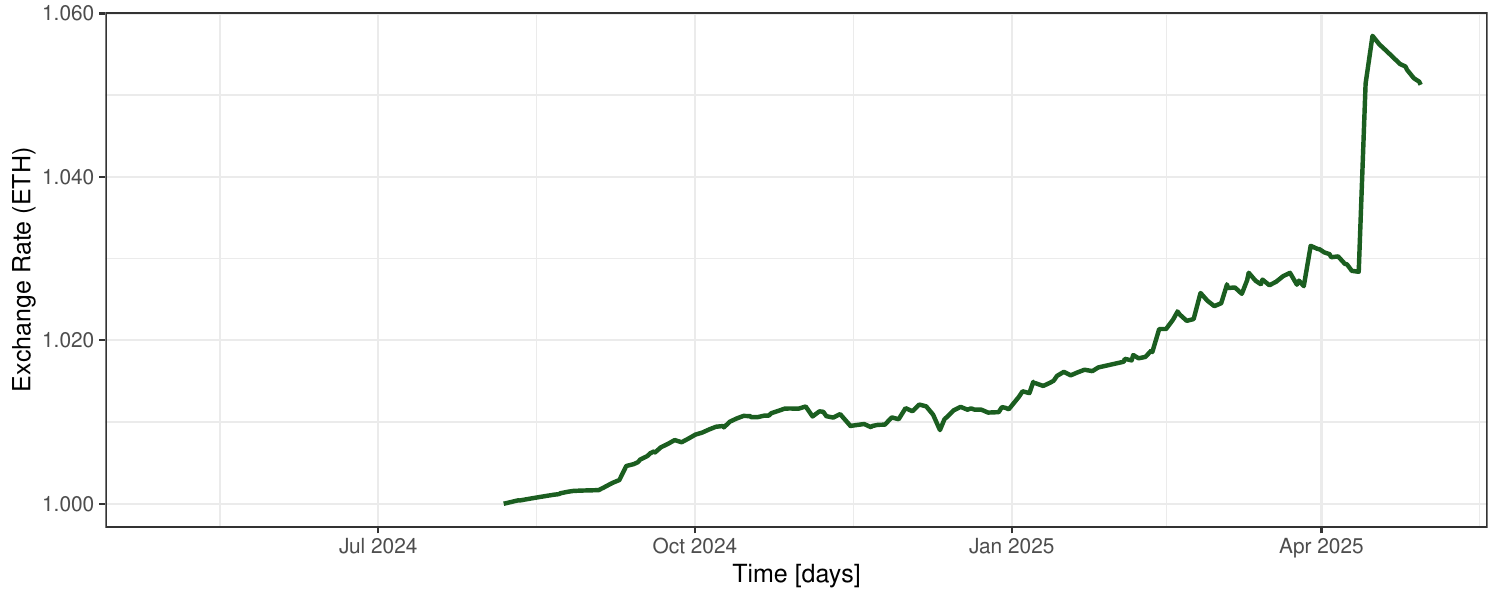}
    \caption{Cian ETH-vault exchange rate: Daily evolution of the share-to-asset exchange rate (ETH--ylstETH) of Cian's ETH vault, which accepts ETH/WETH/stETH/wstETH deposits, over the observation period; anomalous jumps near the end of the window are excluded from the realised-yield computation.}
    \label{fig:cian-exchange-rate}
\end{figure}

Figure~\ref{fig:deposit-investment-comparison} compares the absolute total deposits and total amounts invested into yield strategies for Yearn and Cian over the observation period.

\begin{figure}[h]
\centering
\includegraphics[width=0.8\textwidth]{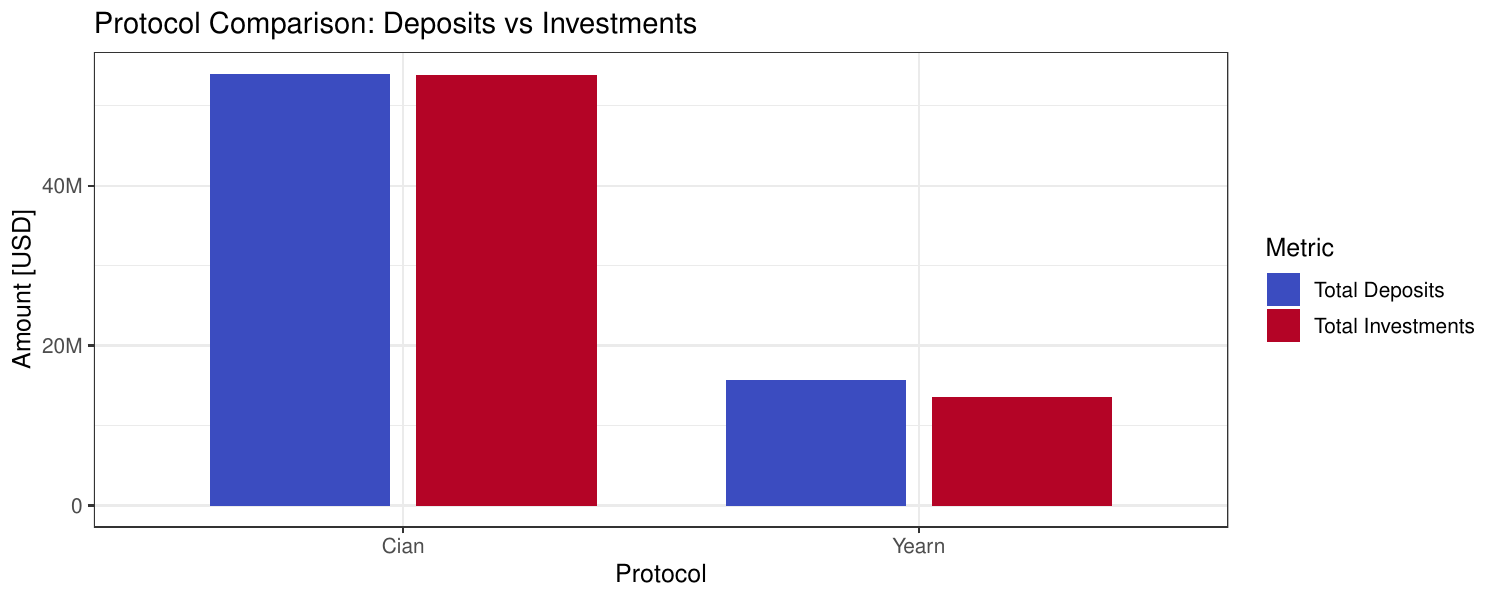}
\caption{Total deposits versus invested capital: Absolute total user deposits and total amounts invested into yield strategies for Yearn and Cian over the observation period, illustrating the share of deposited capital effectively deployed by each aggregator.}
\label{fig:deposit-investment-comparison}
\end{figure}

\fi

\end{document}